\documentclass[twocolumn]{aastex63}

\usepackage{bm}
\usepackage{hyperref}
\usepackage{breakcites}
\usepackage{amsmath,amssymb,amsthm} 
\usepackage{url} 
\usepackage{float}
\usepackage{graphicx}
\usepackage{subfigure}
\usepackage{xcolor}
\usepackage{lineno}
\usepackage{xspace} 
\usepackage{tablefootnote}

\usepackage{chngcntr}
\usepackage{textcomp}

\newcommand{\eqb}{\begin{eqnarray}}
\newcommand{\eqe}{\end{eqnarray}}
\newcommand{\bi}{\begin{itemize}}
\newcommand{\ei}{\end{itemize}}

\newcommand{\txs}{TXS~0506+056 \xspace}
\newcommand{\swift}{\emph{Swift}}
\newcommand{\fermi}{\emph{Fermi}}

\definecolor{royalblue}{RGB}{65,105,225}
\definecolor{royalred}{RGB}{255,40,0}
\shorttitle{Flare Duty Cycle of Gamma-Ray Blazars and Implications for High-Energy Neutrinos}
\shortauthors{Yoshida et al.}
\graphicspath{{./}{figures/}}

\begin{document}

\title{Flare Duty Cycle of Gamma-Ray Blazars and Implications for High-Energy Neutrino Emission}

\correspondingauthor{Kenji Yoshida}
\email{yoshida@shibaura-it.ac.jp}

\author{Kenji Yoshida}
\affiliation{Department of Electronic Information Systems, Shibaura Institute of Technology, 307 Fukasaku, Minuma-ku, Saitama, Japan}

\author{Maria Petropoulou}
\affiliation{National and Kapodistrian University of Athens, 157 72, Athens, Greece}
\affiliation{Institute of Accelerating Systems \& Applications, University Campus Zografos, Athens, Greece}

\author{Kohta Murase}
\affiliation{Department of Physics; Department of Astronomy \& Astrophysics; Center for Multimessenger Astrophysics, Institute for Gravitation and the Cosmos, The Pennsylvania State University, University Park, PA, USA}
\affiliation{School of Natural Sciences, Institute for Advanced Study, Princeton, NJ, USA}
\affiliation{Center for Gravitational Physics and Quantum Information. Yukawa Institute for Theoretical Physics, Kyoto, Kyoto, Japan}

\author{Foteini Oikonomou}
\affiliation{Physics Department, Norwegian University of Science and Technology, 7491, Trondheim, Norway}

\begin{abstract}
Gamma-ray flares of blazars may be accompanied by high-energy neutrinos due to interactions of high-energy cosmic rays in the jet with photons, as suggested by the detection of the high-energy neutrino IceCube-170922A during a major gamma-ray flare from blazar TXS 0506+056 at the $\sim3\sigma$ significance level. 
In this work, we present a statistical study of gamma-ray emission from blazars to constrain the contribution of gamma-ray flares to their neutrino output. We construct weekly binned light curves for 145 gamma-ray bright blazars in the {\it Fermi} Large Area Telescope (LAT) Monitored Source List adding TXS 0506+056. We derive the fraction of time spent in the flaring state (flare duty cycle) and the fraction of energy released during each flare from the light curves with a Bayesian blocks algorithm. 
We find that blazars with lower flare duty cycles and energy fractions are more numerous among our sample. 
We identify a significant difference in flare duty cycles between blazar sub-classes at a significance level of 5~\%.
Then using a general scaling relation for the neutrino and gamma-ray luminosities, $L_{\nu} \propto  (L_{\gamma})^{\gamma}$ with a weighting exponent of ${\gamma} = 1.0 - 2.0$, normalized to the quiescent gamma-ray or X-ray flux of each blazar, we evaluate the neutrino energy flux of each gamma-ray flare. 
The gamma-ray flare distribution indicates that blazar neutrino emission may be dominated by flares for $\gamma\gtrsim1.5$.
The neutrino energy fluxes for one-week and 10-year bins are compared with the declination-dependent IceCube sensitivity to constrain the standard neutrino emission models for gamma-ray flares. Finally, we present the upper-limit contribution of blazar gamma-ray flares to the isotropic diffuse neutrino flux.
\end{abstract}


\section{Introduction} \label{sec:intro}
In 2013 the IceCube Neutrino Observatory discovered an all-sky flux of astrophysical neutrinos with energies from $\sim100$~TeV to several PeV \citep{Aartsen:2013bka,Aartsen:2013jdh}. Since then, the spectrum of this high-energy neutrino flux has been measured to a higher precision with different analyses \citep[e.g.,][]{2015PhRvL.115h1102A, Aartsen2016, Aartsen2020}. The nearly isotropic distribution of the neutrino arrival directions constrains a Galactic contribution to the measured IceCube flux to the level of $\sim10$\% \citep{Albert2018} supporting an extragalactic origin \citep[e.g.,][]{Aartsen2014,Aartsen2015}. 

Many proposals have been put forward to explain the diffuse neutrino flux as the cumulative emission from a population of sources, with several of them even predating its discovery \citep[for recent reviews, see][]{Ahlers2017, Meszaros2017}. Blazars, namely active galactic nuclei (AGN) with relativistic jets closely aligned to our line of sight~\citep{Urry1995}, belong to this long-list of candidate sources. Due to their powerful magnetized jets and copious radiation fields, blazars are promising sites for high-energy neutrino production~\citep[see reviews, e.g.,][and references therein]{2017nacs.book...15M, 2022arXiv220203381M}. 

In 2017 IceCube detected a muon neutrino event with most probable energy of $\sim290$~TeV, IceCube-170922A, in spatial and temporal coincidence with a flaring gamma-ray blazar, TXS 0506+056, at the statistical significance level of $3{\sigma}$ \citep{2018Sci...361.1378I}. Prior to the 2017 flaring event, the IceCube Collaboration also reported $3.5{\sigma}$ evidence for a neutrino excess on top of the atmospheric background expectation from the direction of TXS 0506+056 with an archival search of past IceCube data \citep{2018Sci...361..147I}. 
In addition, some blazars may be powerful enough to produce a strong point-source neutrino signal during flares and have been associated with high-energy neutrinos \citep{Dermer:2014vaa,Kadler16,Petropoulou16,Gao16,Oikonomou19,Paliya20,Petropoulou20,Rodrigues21,Oikonomou21,Sahakyan22}.

Clustering limits \citep{Murase:2016,Capel20} and stacking limits \citep{2017ApJ...835...45A,Neronov17,Hooper19,Smith21,IceCube:2022zbd} suggest that blazars are unlikely to be the dominant sources of the diffuse astrophysical neutrino flux detected by IceCube \citep{Murase:2018,Palladino2019,Yuan2020}. However, they may still appear in stacking analyses of radio-bright blazar arrival directions \citep{Hovatta21,Plavin21,Zhou21,Buson2022}, which may support that the dominant neutrino sources are gamma-ray hidden or opaque \citep{MGA16} or the correlation level may be explained by a handful of the brighest blazars \citep{Murase:2014foa}.
The role of blazar flaring activity in neutrino-source associations is not yet fully understood. Moreover, the results of most stacking analysis are relevant to the time-average blazar emission. 
\cite{Murase:2018} estimated the contribution of blazar flares like the 2017 multi-messenger flare of TXS 0506+056 to the diffuse neutrino flux to be up to $\sim 10\%$ based on the {\it Fermi} All-sky Variability Analysis (FAVA), but the constraint on the contribution from all blazar flares is very uncertain because it is affected by different assumptions \citep{Yuan2020}.

In this paper we expand upon the work of \cite{Murase:2018} by computing the duty cycle and the energy output fraction of gamma-ray ray flares using a much larger sample of blazars observed by the Large Area Telescope (LAT) on board the {\fermi} satellite. To estimate the neutrino flux during gamma-ray flares we use a generic relationship between the all-flavor neutrino and gamma-ray fluxes, i.e. $F^{\rm fl}_{\nu}\propto \left(F_\gamma^{\rm fl}\right)^\gamma$, where $1 \le \gamma \le 2$ \citep[see also][]{Yuan2020}. 
We explore two cases depending on the definition of the neutrino flux in non-flaring periods (i.e. quiescence). 
First, we study the case where the neutrino flux outside flares is benchmarked with the quiescent gamma-ray flux, which has been widely adopted prior to the \txs observations \citep[e.g.,][]{Murase:2014foa,Petropoulou15,Padovani15}. Second, motivated by the modeling studies of \txs we consider a case where the non-flaring neutrino flux is bound by the quiescent X-ray flux. Besides the usual $E_{\nu}^{-2}$ power-law neutrino spectrum, we consider for both scenarios the effects of realistic neutrino spectra, using a spectral template derived for the multi-messenger flare of \txs.
By comparing our neutrino predictions for each blazar in the sample to the time-integrated 10-year IceCube sensitivity and the IceCube sensitivity for transient searches (scaled to one-week) we constrain the aforementioned scenarios. We finally compute the contribution of blazar flares to the all-sky diffuse neutrino flux.

This paper is structured as follows. In Sec.~\ref{sec:analysis} we describe the sample and the \fermi-LAT analysis for generating the light curves. We describe next how we define gamma-ray flaring states using the binned LAT light curves and present our results about the duty cycle of flares and their energy output fraction in Sec.~\ref{sec:flaring_rates}. We discuss the implications of our results on the high-energy neutrino emission from flaring blazars in Sec.~\ref{sec:neutrino_estimates}. We conclude in Sec.~\ref{sec:summary} with a summary of our results. In this work we adopt a flat $\Lambda$CDM cosmology with $H_0=70$~km s$^{-1}$ Mpc$^{-1}$ and $\Omega_{\rm m}=0.30$.

\section{Gamma-ray data and analysis}\label{sec:analysis}
\subsection{Data selection}\label{sec:data}
\begin{figure*}
\begin{center}
\includegraphics[width=0.70\textwidth]{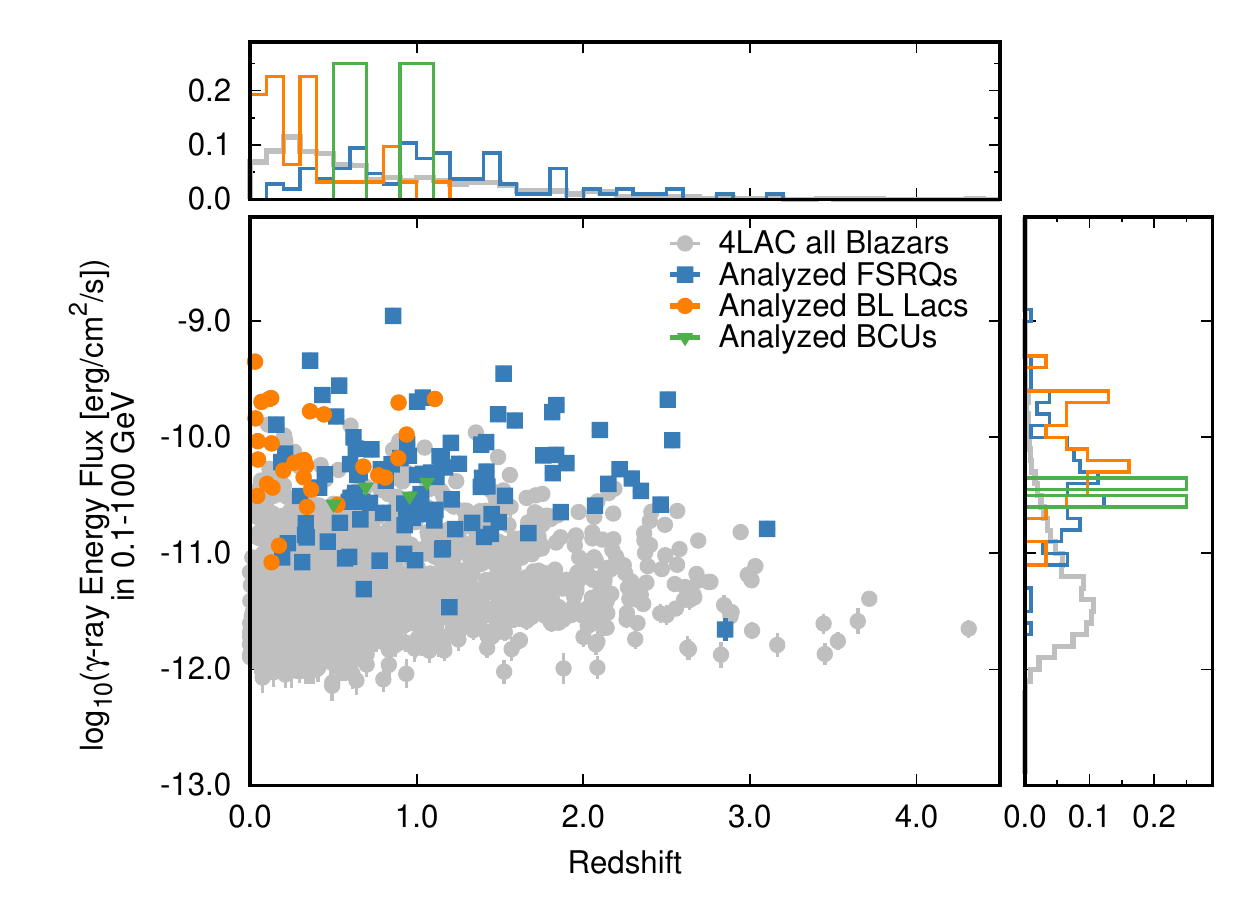}
\end{center}
\caption{Scatter diagram of the time-average gamma-ray energy flux and redshift for the 145 blazars in our sample (106 FSRQs in blue, 31 BL Lacs in orange, and 8 BCUs in green). All blazars in the 4LAC are overlaid (grey markers) for comparison reasons. 
Side panels show the projected histograms on each axis.
The energy fluxes are taken from 4LAC and the redshifts are from 4LAC and SIMBAD \citep{SIMBAD:2019ASPC}.}
\label{fig:z2F_scatthist}   
\end{figure*}

We select 145 gamma-ray bright blazars (Table~\ref{tab:source_list}), consisting of 106 flat spectrum radio quasars (FSRQs), 31 BL Lac objects (BL Lacs), and 8 blazar candidates of uncertain type (BCUs). Among them, the 144 blazars are selected from the \fermi\ Large Area Telescope (LAT) Monitored Source List\footnote{\url{https://fermi.gsfc.nasa.gov/ssc/data/access/lat/msl_lc/}}, which we identified as blazars with the Fourth Catalog of AGN detected by the LAT (4LAC-DR2, hereafter 4LAC) \citep{Fermi4LAC:2020}. The Fermi-LAT team monitors the light curves of a number of bright and transient sources that have flares during the mission to facilitate follow-up multi-wavelength observations of flaring sources. When sources exceed the monitoring flux threshold of $1{\times}10^{-6}$~cm$^{-2}$s$^{-1}$, they are added to the list of monitored sources \footnote{https://fermi.gsfc.nasa.gov/ssc/data/access/lat/msl\_lc/}. Since blazars are known for their intense flares, the list gives the appropriate blazar samples that have flaring fluxes with high statistics. In addition, we added one blazar of TXS~0506+056, whose flaring state is indicated to be in coincidence with IceCube-170922A (IceCube Collaboration et al. 2018a). Our sample comprises the brightest blazars in the 4LAC catalog as shown in Figure~\ref{fig:z2F_scatthist}.
In this figure we show the time-average energy flux in the energy range of 0.1 -- 100~GeV based on 10 years of data as a function of redshift for our sample and all blazars in the 4LAC. 
In this work, we analyze data collected by the LAT for the period 2009--2018 as described in the next section. 

\subsection{Fermi LAT analysis}\label{sec:fermi}
The \fermi\ LAT is a pair conversion telescope sensitive to gamma-rays in the 20 MeV to $>$300 GeV energy range~\citep{Atwood:2009ez}. In this work, we analyze the LAT data from MJD~54682 to MJD~58739 using the {\tt Fermi Science Tools} version 11-05-03\footnote{\url{http://fermi.gsfc.nasa.gov/ssc/data/analysis/software}} and {\tt Fermipy} version 0.17.3 \citep{Fermipy:2017}. 
Following the standard selection recommendation\footnote{\url{https://fermi.gsfc.nasa.gov/ssc/data/analysis/documentation/Cicerone/Cicerone_Data_Exploration/Data_preparation.html}}, we select photons with energies between 100~MeV and 316~GeV that were detected within a radius of $20^{\circ}$ centered on the position of the source of interest, while photons detected with a zenith angle larger than 90$^{\circ}$ with respect to the spacecraft were discarded to reduce the Earth-limb gamma-ray contamination.

The contribution from isotropic and Galactic diffuse backgrounds is modeled using the parameterization provided in the \texttt{iso\_P8R3\_SOURCE\_V2\_v01.txt} and \texttt{gll\_iem\_v07.fits} files, respectively, leaving their all parameters free to vary in the spectral fit. 
Sources in the 4FGL catalog within a radius of $15^{\circ}$ from the source position are included in the model with their spectral parameters fixed to their catalog values \citep{Fermi4FGL:2020}, while the normalization of sources within $3^{\circ}$ is allowed to vary freely during the spectral fit. The spectral fit is performed with a binned likelihood method using the \texttt{P8R3\_SOURCE\_V2} instrument response function with parametrization from 4FGL to characterize the spectral parameters of the source of interest in the 100~MeV--316~GeV energy range. 

The gamma-ray light curve for each source is built with an end-to-end analysis in each time bin using the same processing steps as the spectral analysis. The period between MJD 54682 and MJD 58739 is divided into $\sim$570 7 day-long bins. A model spectral fit\footnote{The spectral models used are from the 4FGL, e.g., PowerLaw, LogParabola, and PLSuperExpCutoff. In LogParabola the free parameters are norms, $\alpha$, and $\beta$.} is performed in each time bin with the free model parameters. 
The best-fit parameters are then used to calculate the gamma-ray flux in the 0.1--316~GeV energy range for each time bin of the light curve. 

\begin{figure*}
\gridline{\fig{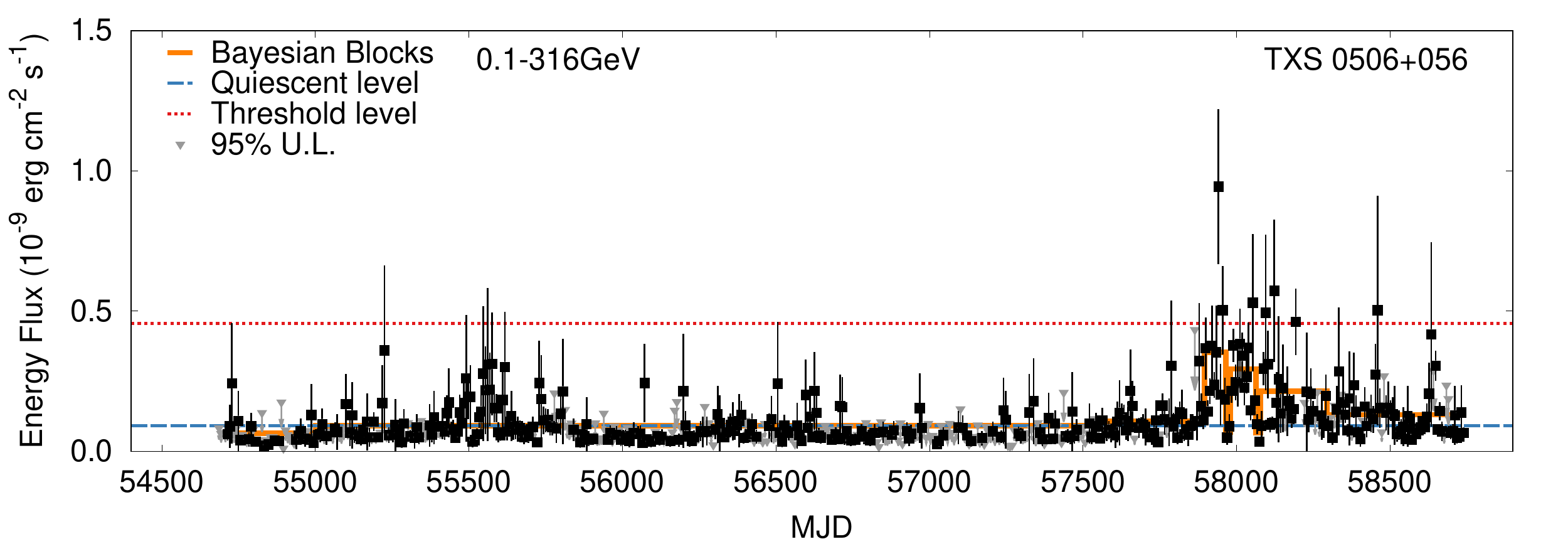}{0.49\textwidth}{(a) TXS 0506+056}
          \fig{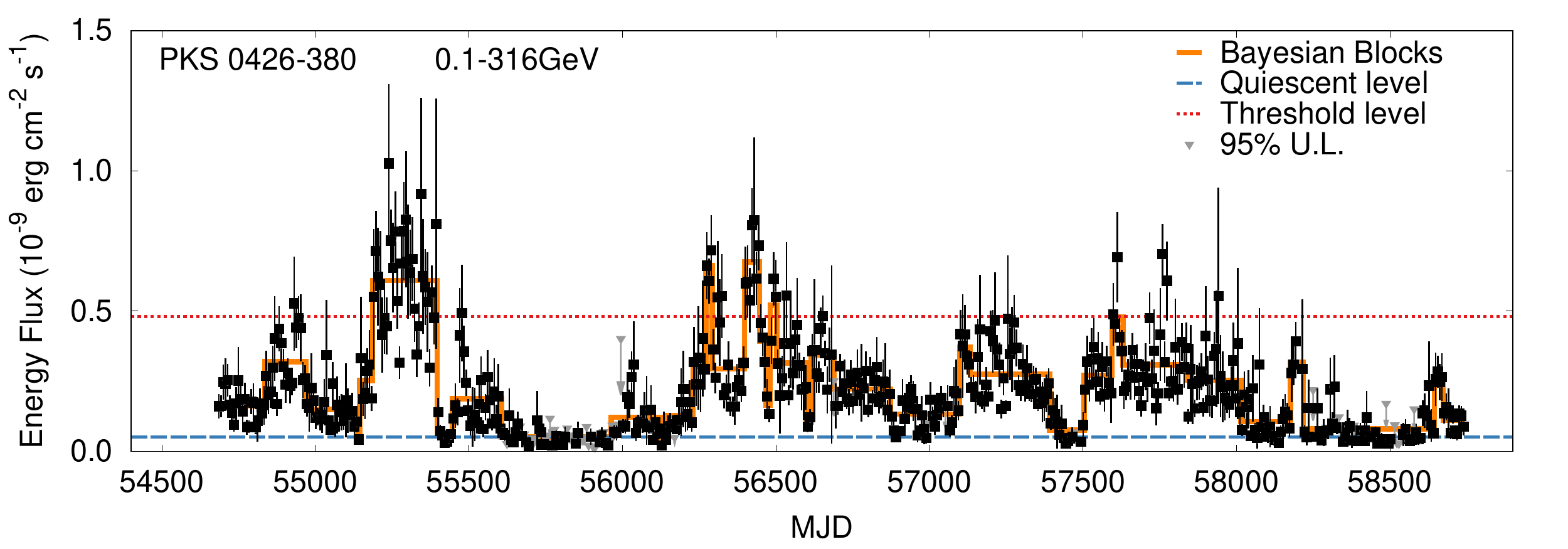}{0.49\textwidth}{(b) PKS 0426-380}}
\gridline{\fig{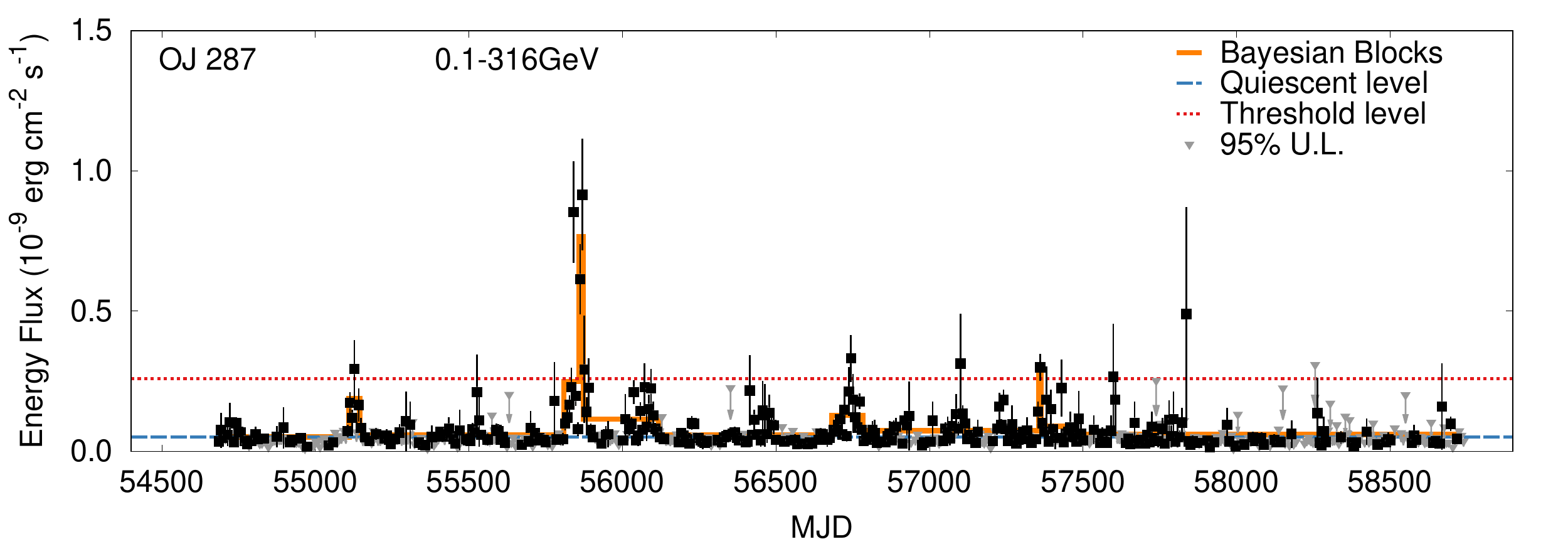}{0.49\textwidth}{(c) OJ 287}
          \fig{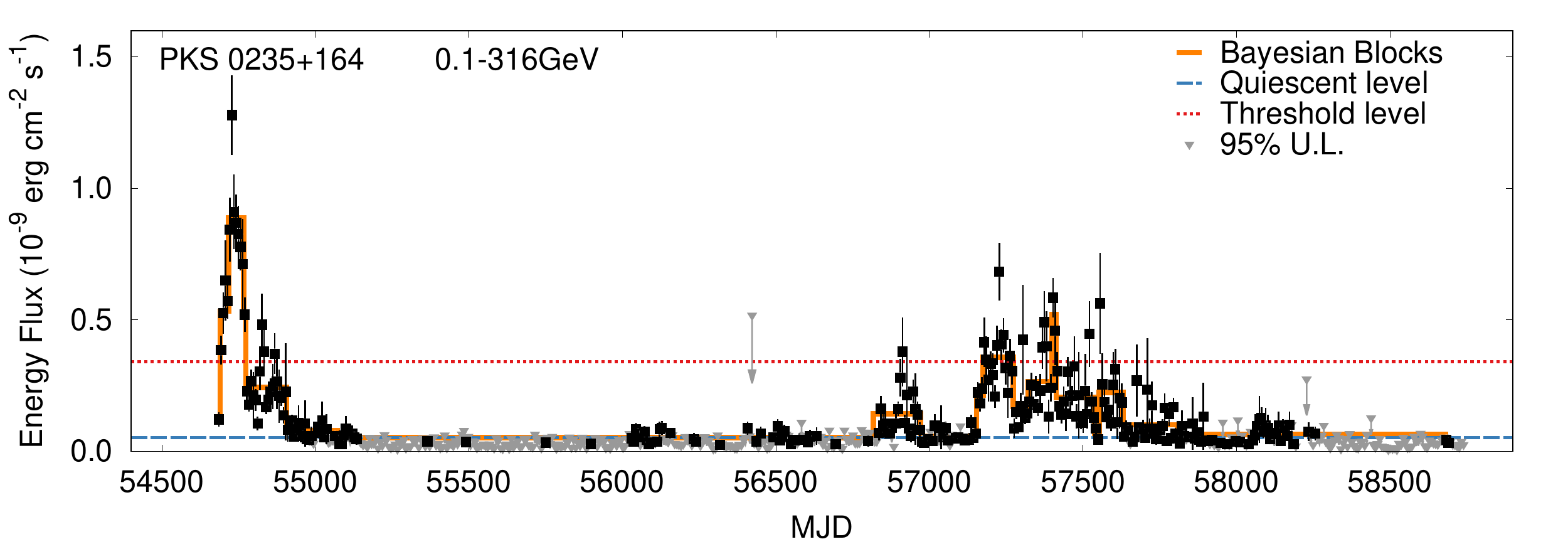}{0.49\textwidth}{(d) PKS 0235+164}
          }
\caption{
Indicative examples of weekly-binned blazar gamma-ray light curves (black points) in the 0.1-316~GeV energy range. Grey triangles indicate 95\% upper limits.
The quiescent flux and flare threshold levels are plotted with dashed blue and red lines, respectively (for a definition of both flux thresholds, see Sec.~\ref{sec:flares}). 
The Bayesian blocks representation of each light curve is overplotted with orange solid lines.  
The 1-316~GeV light curves of the same sources can be found in Appendix~\ref{sec:lcHE}.
\label{fig:lc}}
\end{figure*}

Figure~\ref{fig:lc} shows examples of gamma-ray light curves for four blazars, including the first plausible cosmic neutrino source, \txs. In all light curves, we show flux points with their $1\sigma$ uncertainties (black symbols), if the source is detected with a test statistic (TS) larger than 9 (corresponding to a $3\sigma$ excess) and the flux in one bin is larger than its uncertainty. Otherwise, we show the flux upper limits at the 95~\% confidence level.

\section{Gamma-Ray flare distributions}\label{sec:flaring_rates}
In this section, we start by defining gamma-ray flaring states using the weekly-binned \fermi\ LAT light curves (Section~\ref{sec:flares}). We then derive the flare duty cycle (i.e., fraction of time spent in flaring states) and the flare energy fraction (i.e., fraction of energy released in flaring states) for all sources in our sample, and discuss potential differences between BL Lac objects and FSRQs (Section~\ref{sec:duty_cycle}). A comparison of our results with those reported by \cite{Murase:2018} can be found in Appendix~\ref{sec:fava}.

\subsection{Definition of flares}\label{sec:flares}
\subsubsection{Quiescent flux level} \label{sec:quiescent}
Although there is no rigorous way to define the quiescent flux level of blazar emission, it is reasonable to assume that the quiescent state has a relatively low and stable flux. In this study, we adopt the following procedure to derive the quiescent gamma-ray flux level of a blazar: 
\begin{enumerate}
\item We apply a Bayesian blocks algorithm to each gamma-ray light curve with a false alarm probability $p_{0} = 0.05$, adopting the so-called point measurements fitness function for the Bayesian blocks algorithm 
\citep{Scargle:2013, Astropy:2013, Astropy:2018}. 

\item The Bayesian blocks algorithm provides the partitions of the light curve into blocks where the flux is regarded to be constant. 
This is an optimal step-function representation of the light curve.

\item We consider the minimum average flux as a representative of the source quiescent flux, under the following condition: the number of data points within the selected block is more than or equal to a half of the average number of data points of all other blocks. 
This condition prevents us from using an insufficient amount of data points to estimate the quiescent flux level. We note that the quiescent flux levels hardly depend on the choice of $p_{0}$ (in the range $0.01 - 0.1$) as described in Sec.~\ref{sec:threshold_level}.
\end{enumerate}

Figure~\ref{fig:lc} shows indicative examples of weekly-binned \fermi\ LAT light curves from our sample (black points) and the Bayesian blocks representations (solid orange lines). The quiescent flux levels derived using the method described above are also indicated with dashed cyan lines. 
To verify the quiescent flux level being identified properly, we derived the gamma-ray quiescent fluxes by using coarser time binning for the light curves. See Appendix~\ref{sec:time_binning} for details.

\begin{figure*}
\gridline{\fig{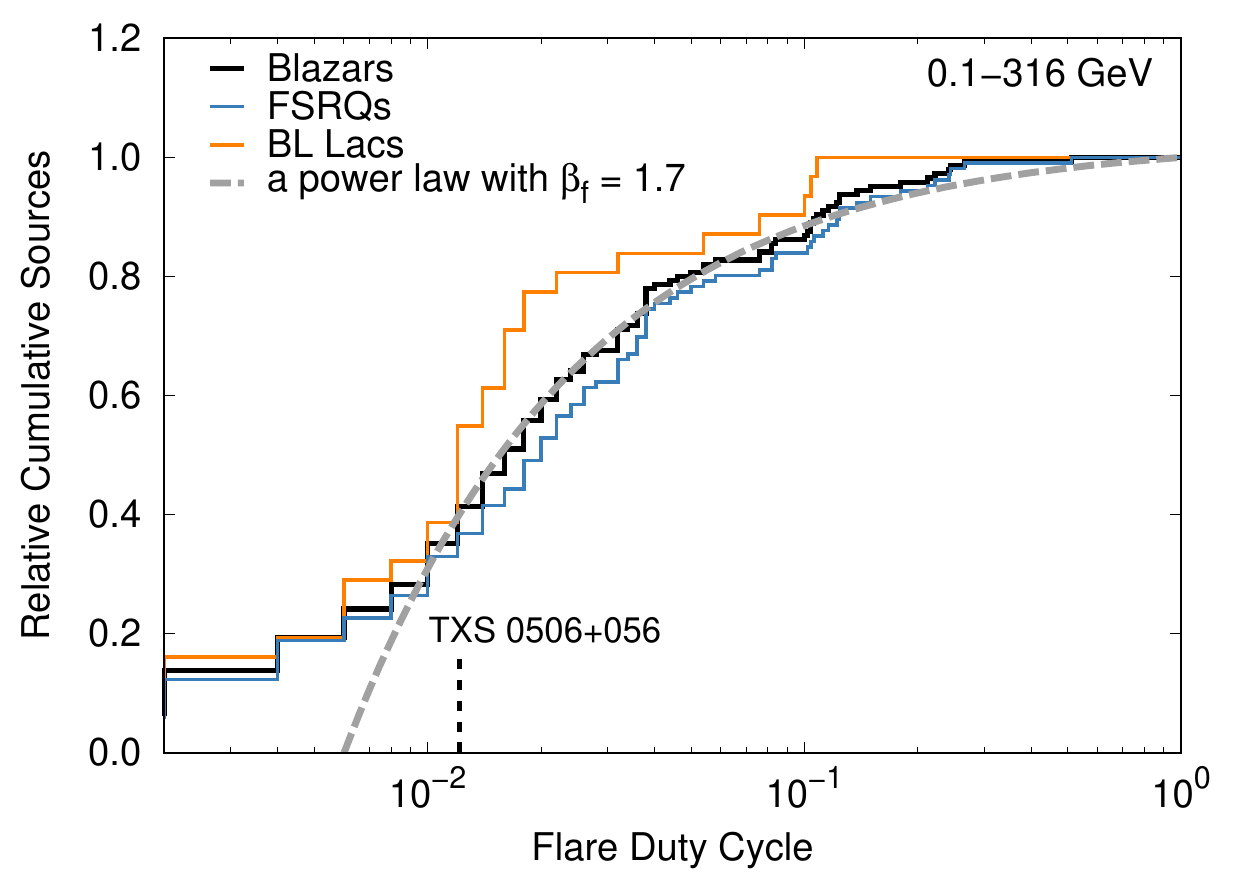}{0.45\textwidth}{(a)}
          \fig{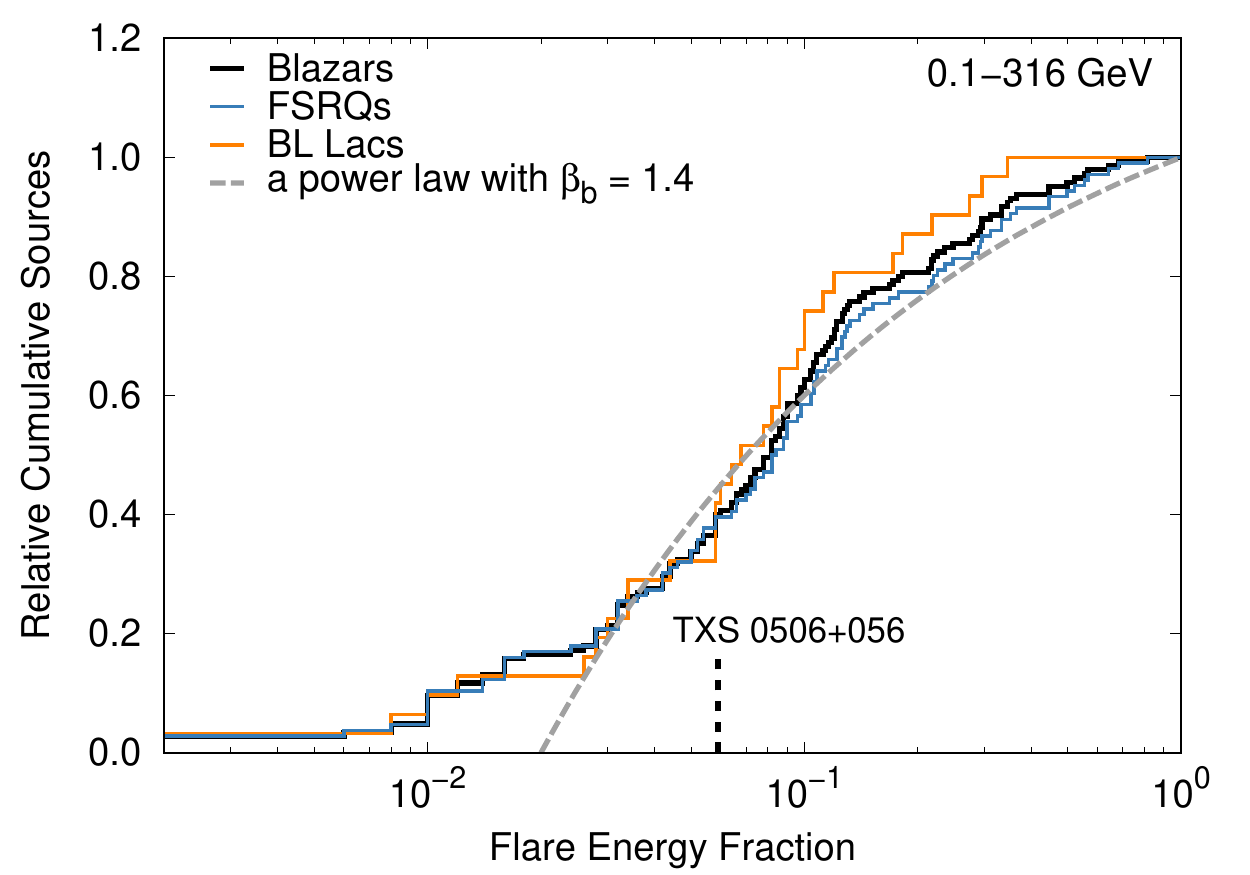}{0.45\textwidth}{(b)}
          }
\caption{
Cumulative histograms (normalized to one) of the $0.1-316$~GeV gamma-ray flare duty cycles (a) and flare energy fractions (b) of 145 blazars (black solid histogram), 106 FSRQs (blue solid histogram), and 31 BL Lacs (orange solid histogram). In both panels, the vertical dashed line indicates the values for \txs.The dashed gray lines depict cumulative power-law functions (see text for more details).}
\label{fig:df_bf_cumula}
\end{figure*}

\subsubsection{Flaring threshold level} \label{sec:threshold_level}
Various definitions of flaring states in blazar emission exist in the literature \citep[e.g.,][]{Resconi:2009, Nalewajko:2013,Meyer:2019, 2022MNRAS.510.4063S}. Here, we define as a gamma-ray flare, any flux point 
of the weekly-binned light curve that exceeds a certain threshold, $F_\gamma^{\rm th}$, given by 
\begin{equation}
\label{eq:threshold}
F_\gamma^{\rm th} = F_\gamma^{q} + s {\langle}F_\gamma^{\rm err}{\rangle}, 
\end{equation}
where 
$F_\gamma^{q}$ is the gamma-ray quiescent flux level, ${\langle}F_\gamma^{\rm err}{\rangle}$ is the average uncertainty of the gamma-ray fluxes\footnote{${\langle}F_\gamma^{\rm err}{\rangle} =(\sum_{i=1}^{N} \sigma_{i})/N$, where $N$ is the number of time bins of a light curve and $\sigma_{i}$ is the flux uncertainty of the $i$-th time bin.},  and $s$ corresponds to the significance above the quiescent flux level in units of the standard deviation ${\sigma}$. Unless otherwise noted, we use $s = 6$ in this work. 
The flaring threshold levels are indicated with dotted red lines in Figure~\ref{fig:lc}.

\subsection{Flare duty cycle and flare energy fraction}\label{sec:duty_cycle}
The fraction of time spent in the flaring state, i.e., the flare duty cycle, is defined as 
\begin{equation}
f_{\rm fl} = \frac{1}{T_{\rm tot}} \int_{F_\gamma^{\rm th}} dF_{\gamma}\frac{dT}{dF_\gamma}
\label{eq:duty_cycle}
\end{equation}
where $T_{\rm tot}$ is the total observation time,  $F_\gamma$ is the gamma-ray energy flux, and $T$ is the time spent at the respective flux level.
The fraction of energy emitted in the flaring state, i.e., the gamma-ray flare energy fraction, is given by~\citep{Murase:2018}
\begin{equation}
b^{\gamma}_{\rm fl} = \frac{1}{F_\gamma^{\rm ave} T_{\rm tot}} \int_{F_\gamma^{\rm th}} 
dF_{\gamma}F_{\gamma}\frac{dT}{dF_\gamma}
\label{eq:gamma_bfl}
\end{equation}
where $F_\gamma^{\rm ave}$ is the average gamma-ray energy flux over the whole observation period. 
Note that \cite{Murase:2018} introduced the flare duty cycle and energy fraction with gamma-ray luminosity $L_{\gamma}$.

\begin{figure*}
\begin{center}
\includegraphics[width=0.60\textwidth]{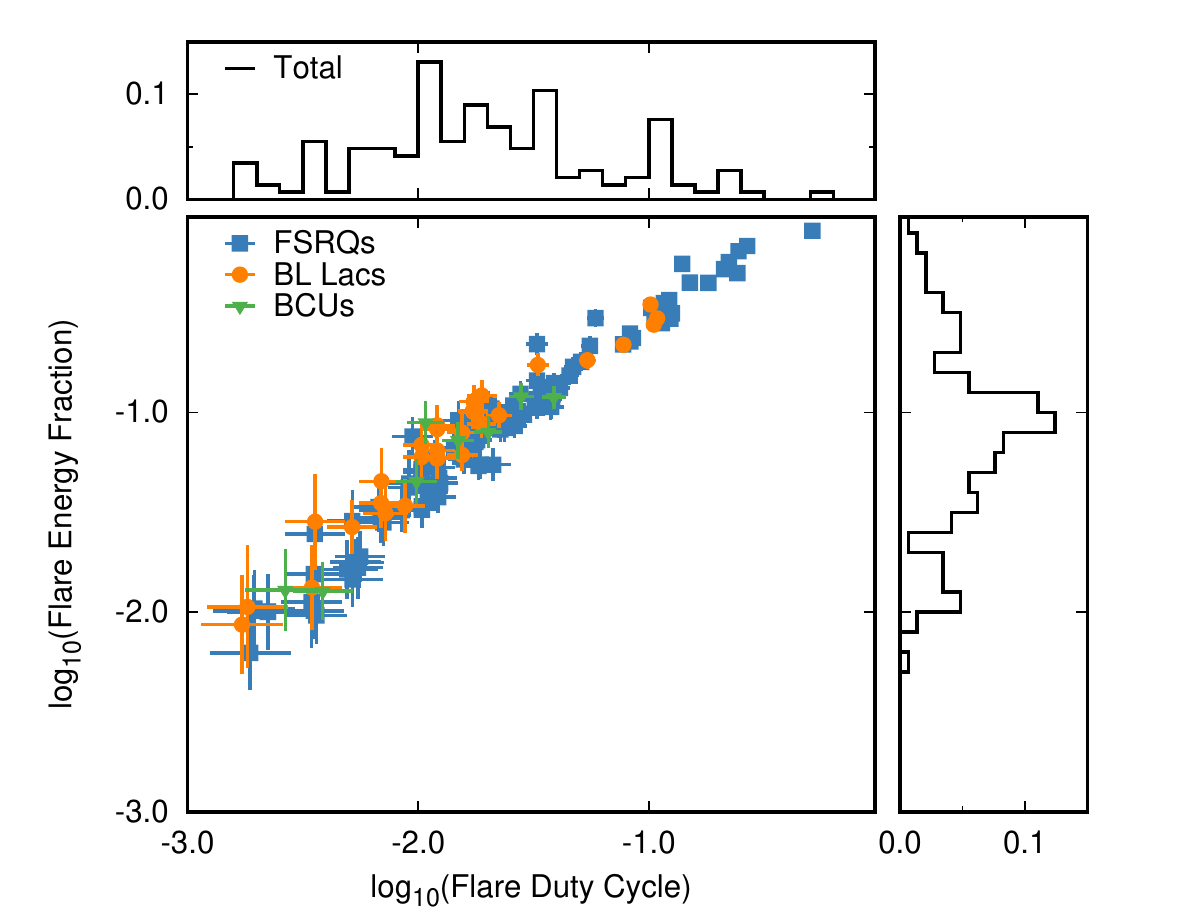}
\end{center}
\caption{
Scatter plot of the $0.1-316$~GeV gamma-ray flare duty cycles and flare energy fractions of 141 blazars (103 FSRQs in blue squares, 30 BL Lacs in orange circles, and 8 BCUs in green triangles). 
Side panels show the projected histograms on each axis. 
}
\label{fig:df2bf_scatthist}
\end{figure*}

Figure~\ref{fig:df_bf_cumula} shows the cumulative distributions of the $0.1-316$~GeV gamma-ray flare duty cycles ($f_{\rm fl}$) and flare energy fractions ($b^{\gamma}_{\rm fl}$) of 145 blazars (106 FSRQs, 31 BL Lacs, and 8 BCUs). 
The $f_{\rm fl}$ and $b^{\gamma}_{\rm fl}$ values for \txs~are indicated with vertical dashed lines. 
\txs~belongs to the 30~\% (40~\%) of the cumulative distribution of the flare duty cycle (flare energy fraction).
Thus, \txs~is not an extraordinary source in terms of its gamma-ray flaring activity.
We also perform a two-sample Kolmogorov-Smirnov test to check if the cumulative distributions of either $f_{\rm fl}$ or $b^\gamma_{\rm fl}$ are the same for FSRQs and BL~Lacs. 
Under the null hypothesis that the FSRQ and BL Lac distributions are derived from the same parent distribution, the probabilities are $p = 0.039$ for the flare duty cycles distributions with the K-S statistic $D = 0.279$, and $p = 0.521$ for the gamma-ray flare energy fraction distributions with $D = 0.159$, respectively.
Hence, we can reject the null hypothesis that the distributions of flare duty cycles for FSRQs and BL Lacs are drawn from the same parent population at a significance level of 5~\%. 
On the other hand, we cannot exclude the null hypothesis for the flare energy fractions of FSRQs and BL Lacs at a significance level of 10~\%.

Figure~\ref{fig:df2bf_scatthist} shows a strong correlation between the flare duty cycles ($f_{\rm fl}$) and the flare energy fractions ($b^\gamma_{\rm fl}$) of 141 blazars (103 FSRQs, 30 BL Lacs, and 8 BCUs); here, non-flaring sources are excluded at the $6\sigma$ threshold.
We find an almost linear relation between the two quantities for $f_{\rm fl}\lesssim 0.1$, which however becomes sublinear for higher flaring duty cycles. The side panels of Figure~\ref{fig:df2bf_scatthist} show histograms of the logarithm of the flare duty cycle and energy fraction. 
It can be shown that both distributions are well represented by power-law functions, namely $dN/df_{\rm fl} \propto f_{\rm fl}^{-\beta_f}$ for $f_{\rm fl}\gtrsim 0.01$ and $dN/db_{\rm fl} \propto (b_{\rm fl}^{\gamma})^{-\beta_b}$ for $b^\gamma_{\rm fl}\gtrsim 0.03$, with $\beta_{f} ~\sim 1.7$ and $\beta_b ~\sim 1.4$.
In other words, blazars with lower flare duty cycles and flare energy fractions are more common in the sample compared to sources that are commonly flaring in gamma-rays. 
Figure~\ref{fig:df_bf_cumula} also depicts the cumulative power-law functions with ${\beta}_f = 1.7$ for $0.006-1.0$ flare duty cycles and ${\beta}_{b} = 1.4$ for $0.02-1.0$ flare energy fractions.

\begin{figure*}
\gridline{\fig{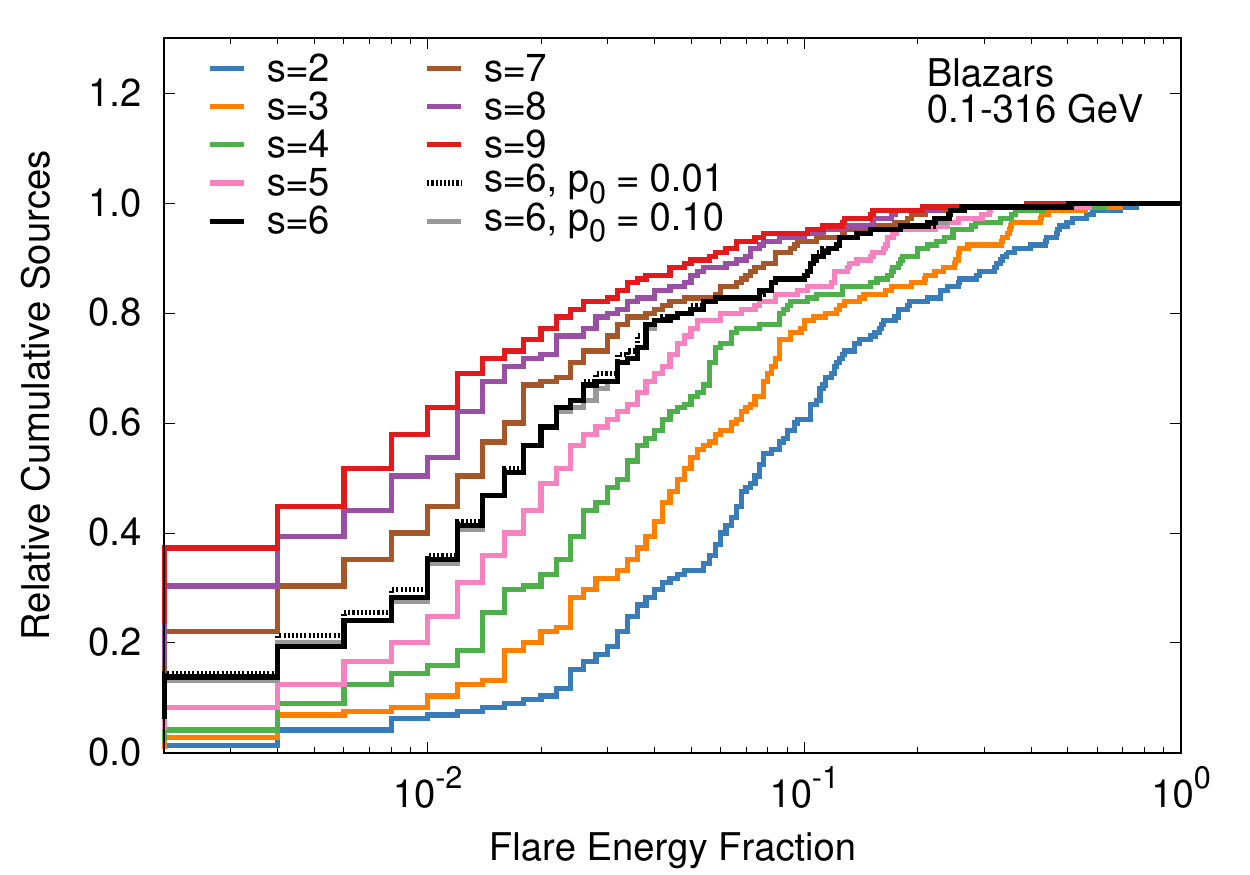}{0.45\textwidth}{(a)}
         \fig{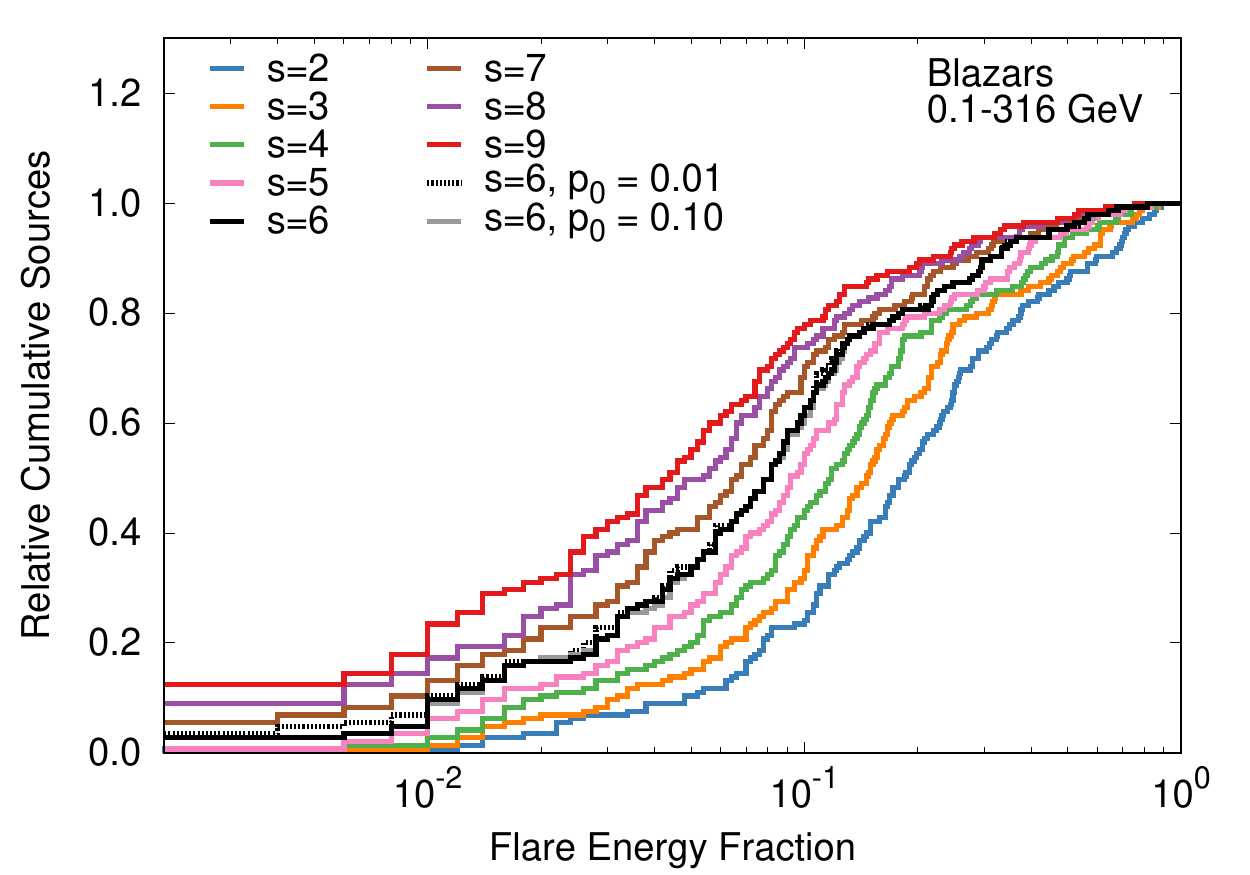}{0.45\textwidth}{(b)}
         }
\caption{
Cumulative histograms of the $0.1-316$~GeV gamma-ray flare duty cycles (a) and flare energy fractions (b) of 145 blazars for different integer values of $s$ ranging from 2 to 9. In the case of $s = 6$ cumulative histograms of $p_{0} = 0.01, 0.10$ are also plotted.}
\label{fig:sig2df_bf_cumula}
\end{figure*}

So far, we have presented results for $f_{\rm fl}$ and $b^\gamma_{\rm fl}$ obtained using the $6\sigma$ threshold for the definition of flares. To check how our results depend on the choice for $s$, we construct the cumulative histograms of both quantities for integer values of $s$ ranging from 2 to 9. Our results are shown in Figure~\ref{fig:sig2df_bf_cumula}. 
By lowering the threshold for the flaring flux, both $f_{\rm fl}$ and $b^\gamma_{\rm fl}$ distributions, shift to higher values as expected. The duty cycle distribution tend to saturate only for $s\ge 8$. Still, for $s\ge 6$, we find small differences in the cumulative distributions of the duty cycle for $f_{\rm fl} \gtrsim 0.05$. This suggests that our default choice of $s=6$ is sufficient for describing the high-end of the duty cycle distribution.  The impact on the flare energy fraction distribution is smaller, as indicated by the smaller spread of the curves. For completeness, we also show the cumulative histograms for $s=6$ and two choices of the false alarm probability ($p_{0} = 0.01$ and $0.10$) of the Bayesian block algorithm that we applied to the \fermi-LAT light curves (see Section~\ref{sec:quiescent}). 
The distributions are not sensitive to the choice of $p_0$ at least for the values we consider. 

\begin{figure}
  \plotone{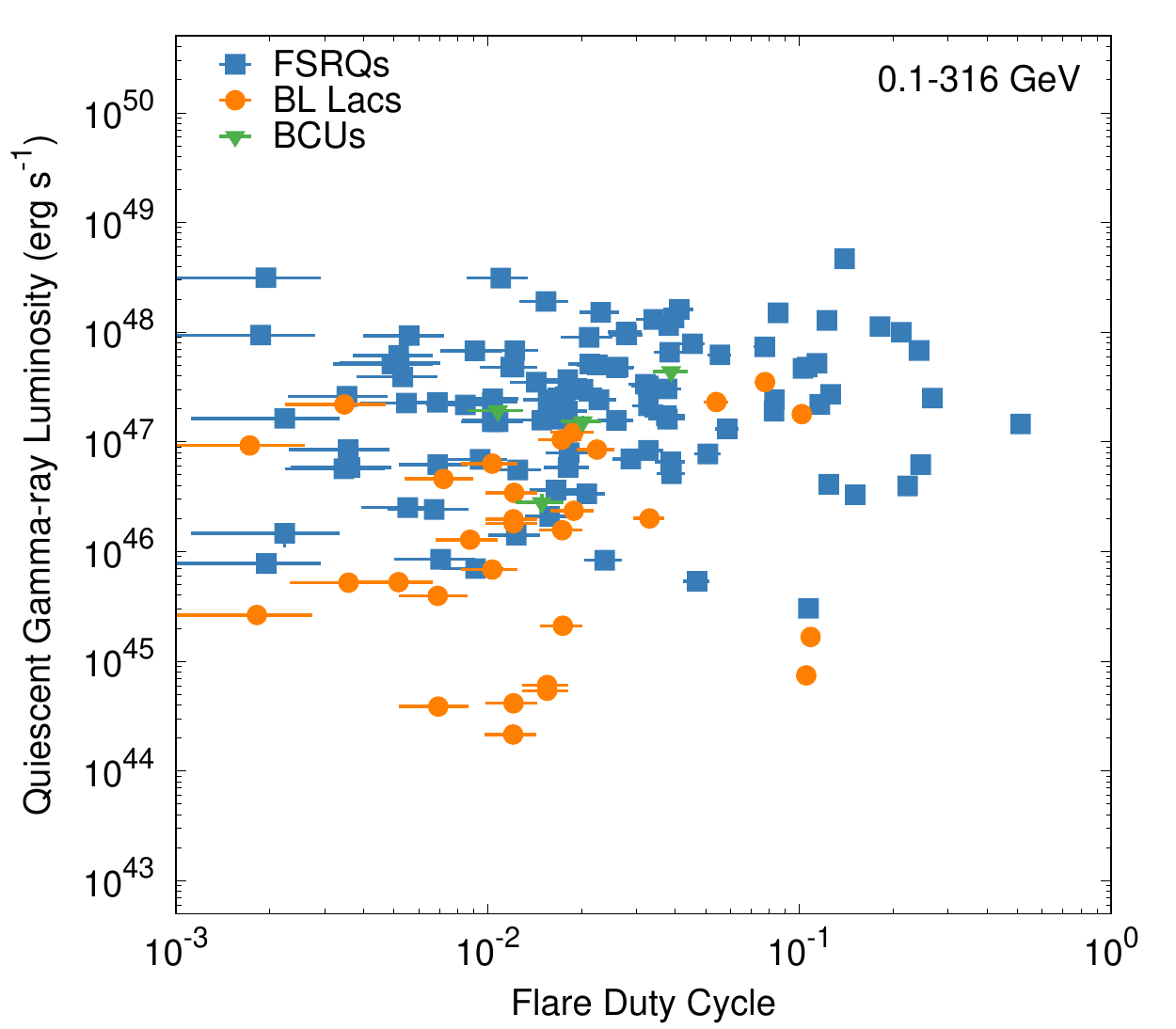}
  \caption{Scatter plot between $0.1-316$~GeV gamma-ray flare duty cycle and quiescent gamma-ray luminosity for 137 blazars of 103 FSRQs (blue squares), 30 BL Lacs (orange circles), and 4 BCUs (green triangles).
  }
  \label{fig:gfl_df2L}
\end{figure}

Having determined the flare duty cycle for different blazar subclasses and having discussed the effects of the flare threshold value, we can now check if there is a correlation between the flare duty cycle and gamma-ray luminosity. 
In other words, we would like to check if the (quiescent) gamma-ray luminosity is a good indicator of the flaring activity of a blazar. 
We compute the quiescent gamma-ray luminosity as $L_{\gamma}^{q} = 4{\pi}d_{L}^{2} F_\gamma^{q}$ without the correction for extragalactic background light absorption, where $d_{L}$ is the luminosity distance calculated with the redshift $z$ and $F_\gamma^{q}$ is the $0.1-316$~GeV gamma-ray quiescent flux (see  Section~\ref{sec:quiescent}). According to the 4FGL catalog, the blazars for $z > 1.0$ show spectral turnovers less than 100~GeV, indicating that the extragalactic background light absorption is irrelevant.
As an uncertainty in the flare duty cycle we use the square root of the sum of squares of the flaring time-bin errors, which are half of the time-bin width, over the total observation time.
As shown in Figure~\ref{fig:gfl_df2L}, there is no correlation between the gamma-ray flare duty cycle and the quiescent gamma-ray luminosity for neither blazar subclass. Similar results are found for different flare threshold values $s = 2-9$.

\section{Implications for high-energy neutrino emission}\label{sec:neutrino_estimates} 
\begin{figure}
\begin{center}
\includegraphics[width=0.45\textwidth]{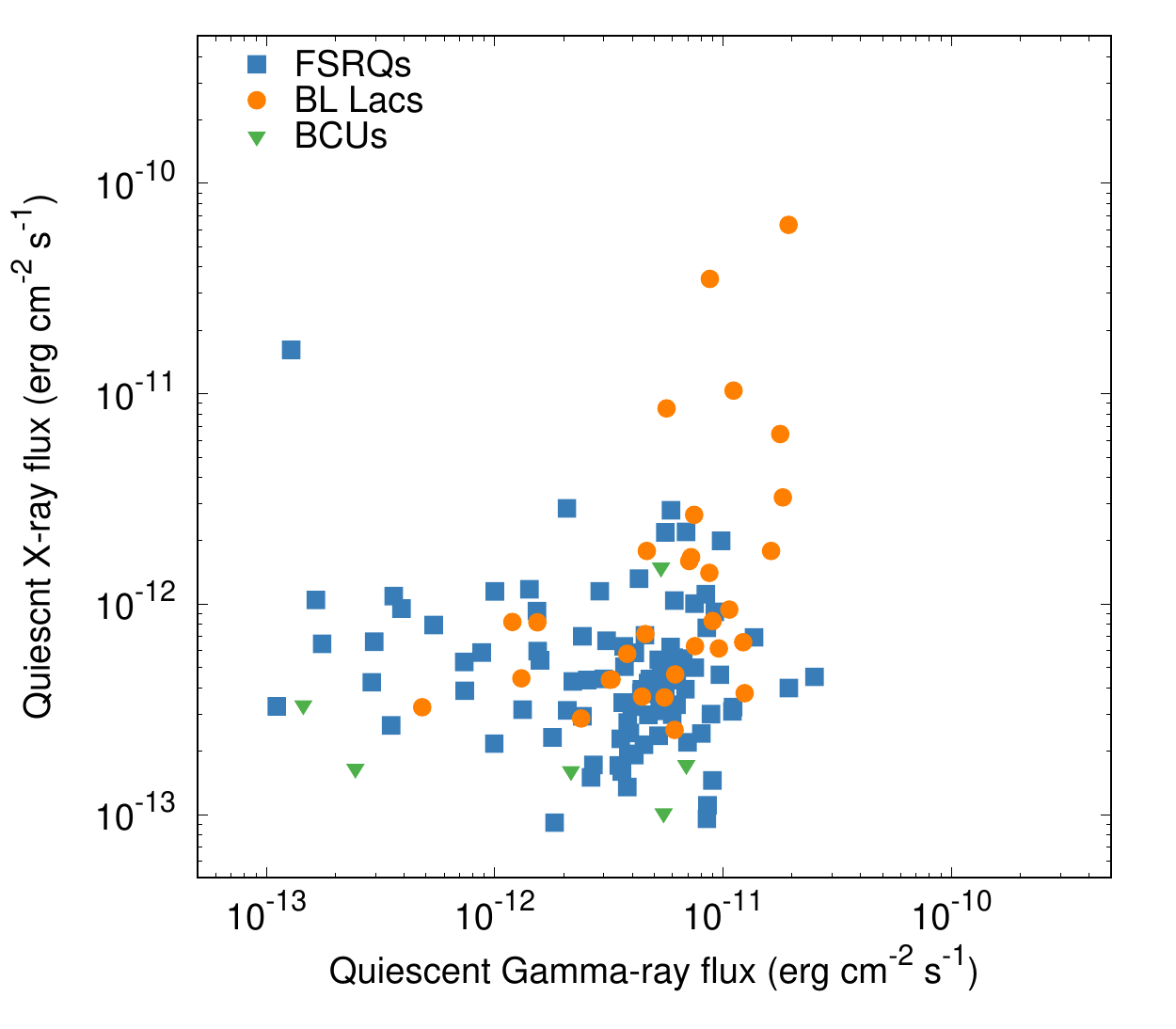}
\end{center}
\caption{Scatter plot between the quiescent gamma-ray and X-ray fluxes for 138 blazars of 101 FSRQs (blue squares), 31 BL Lacs (orange circles), and 6 BCUs (green triangles). The isolated upper left FSRQ is 3C 273. }
\label{fig:gq2xq_scatt}
\end{figure}

In order to discuss implications for neutrino emission, we need to know the physical relationship between neutrinos and gamma rays. In $pp$ scenarios for either gamma-ray or neutrino data, neutrinos and gamma rays can be simply connected via the hadronic production process itself~\citep{Murase2013}. 
However, blazar gamma-ray emission is usually explained by primary leptonic emission (e.g., inverse Compton scattering) and/or secondary cascade emission from photohadronic ($p\gamma$) interactions, thus making the multimessenger connection much more model dependent.
It is therefore useful to parameterize the multimessenger relationship in a generic manner using the luminosity weighting factor $\gamma$ \citep[e.g.,][]{Yuan2020},  
\eqb
    L^{\rm fl}_{\nu} = 
    L^{q}_{\nu}\left(\frac{L^{\rm fl}_{\gamma}}{L^{q}_{\gamma}}\right)^\gamma,
\eqe
where $L^{\rm fl}_{\nu}$ is the flare neutrino luminosity, $L^{q}_{\nu}$ is the quiescent neutrino luminosity, and $L^{\rm fl}_{\gamma}$ is the gamma-ray luminosity in the flaring state. Theoretical models typically predict $\gamma \sim 1.0-2.0$ \citep[e.g.,][]{Murase:2018}.
Note that the differential gamma-ray energy flux, $F_{E_\gamma}$, is related to the bolometric gamma-ray luminosity as 
\eqb
    L_\gamma = 4\pi d_L^2 \int d E_\gamma \, F_{E_\gamma}. 
\eqe
There are different ways to estimate the non-flaring (quiescent) neutrino flux, $F^{q}_{\nu}$. In what follows, we explore two cases where $F^{q}_{\nu}$ is benchmarked with the quiescent gamma-ray flux (scenario 1) or quiescent X-ray flux (scenario 2). For each scenario, we consider the impacts of the neutrino spectrum. 
For illustration purposes, we present in Figure~\ref{fig:gq2xq_scatt} a scatter plot of the quiescent gamma-ray and X-ray fluxes for 138 blazars of 101 FSRQs, 31 BL Lacs, and 6 BCUs.
Note that the majority of sources has $F_{\gamma}^{q}>F_{\rm X}^{q}$, which relates also to their broadband spectral types.

\subsection{Gamma-ray proxy (scenario 1)}
One of the ways to scale the neutrino luminosity is through the gamma-ray luminosity. Such a scaling is reasonable in hadronic models for gamma-ray emission, but it is non-trivial in leptonic models. Nevertheless, it may approximately hold given that the Compton dominance parameter is similar. 
The quiescent muon neutrino flux is benchmarked with the quiescent gamma-ray flux, and the flaring muon neutrino flux is written as
\eqb    
    E_{{\nu}_{\mu}}F^{\rm fl}_{E_{{\nu}_{\mu}}} = E_{{\nu}_{\mu}}F^{q}_{E_{{\nu}_{\mu}}} \left(\frac{F^{\rm fl}_{\gamma}}{F^{q}_{\gamma} }\right)^\gamma
    = A_{\gamma} \, \frac{E_\gamma F^{q}_{E_\gamma}}{3} \left( \frac{F^{\rm fl}_{\gamma}}{F^{q}_{\gamma} }\right)^\gamma , 
    \label{eq:scenarioG} 
\eqe
where $A_{\gamma}$ is a normalization parameter, and the factor of 3 comes from neutrino flavour mixing in vacuum ${\nu}_{e} : {\nu}_{\mu} : {\nu}_{\tau} ~{\simeq}~ 1:1:1$.
Here we relate the differential quiescent neutrino flux at 300 TeV to the quiescent gamma-ray flux at the pivot energies for each weakly time bin. The pivot energies, which are chosen close to the decorrelation energy to minimize the correlation between the fitted gamma-ray spectral parameters, range from 0.1~GeV to 316~GeV even in one source. The typical values are, however, $0.2 - 0.3$~GeV. Then, the quiescent gamma-ray flux $E_\gamma F^{q}_{E_\gamma}$ is estimated from the $E_\gamma F_{E_\gamma}$ light curve at the pivot energies in the same way as the quiescent gamma-ray flux -- see Section \ref{sec:quiescent}.

In photomeson production interactions in the blazar jet, neutrinos and other secondaries produced via the neutral and charged pion decay chains carry comparable energy fluxes (see e.g., equations (13) and (14) of~\citealt{Murase:2018}). As a result, the neutrino energy flux is accompanied by a comparable energy flux of electromagnetic cascade emission at lower energies.  However, not all X-rays and gamma-rays are produced in photomeson production interactions. They may as well originate from Bethe-Heitler pair production and leptonic processes that are not accompanied by neutrinos. Therefore, we adopt $A_{\gamma} = 1$ as an upper limit for the normalisation parameter. A stronger upper limit on $A_{\gamma}$ can be obtained from the non-detection of neutrinos from some of the blazars in our sample as illustrated in the following figures.

Figure~\ref{fig:sindecl2nufl_gqb_A1g15} shows the estimated muon neutrino flare fluxes from scenario 1 with $A_\gamma = 1.0$ and ${\gamma}=1.5$ as a function of ${\sin}({\delta})$, where $\delta$ is the source declination,  neglecting the muon neutrino flux equal to or less than its uncertainty. Results for one-week and 10-year bins are shown in the left and right panels, respectively\footnote{In Appendix F, we also present the estimated muon neutrino flare fluxes using the gamma-ray photon fluxes instead of the gamma-ray energy fluxes.}. The neutrino flare fluxes for a 10-year bin are the averages of the weekly neutrino flare fluxes, with zero fluxes assumed for the non-flaring states. For both binning choices there are several blazar flares with predicted neutrino fluxes that exceed the IceCube sensitivity. The lack of neutrino detections from these sources can therefore be used to derive limits on the normalization $A_\gamma$ for different values of the index $\gamma$ (gray lines), as demonstrated in Figure~\ref{fig:g2Ag}. As expected, the weaker the relation between neutrino and gamma-ray energy fluxes is (i.e., lower $\gamma$ values), the higher $A_\gamma$ can be. 
The red, orange, brown, and purple lines indicate stringent limits obtained for a couple of sources in our sample, such as 3C~454.3, 3C~279, PKS~1502+106, and PKS 1510-089. 
The estimated neutrino flare fluxes with a 10-year bin are lower than those with a one-week bin because of the averaging. 
Meanwhile, the IceCube 90\% sensitivity for a 10-year bin is much lower than that for one-week bin. Hence, the limits placed on $A_{\gamma}$ from the 10-year binning analysis are more stringent than those set by the weekly binned analysis.

\begin{figure*}
\gridline{\fig{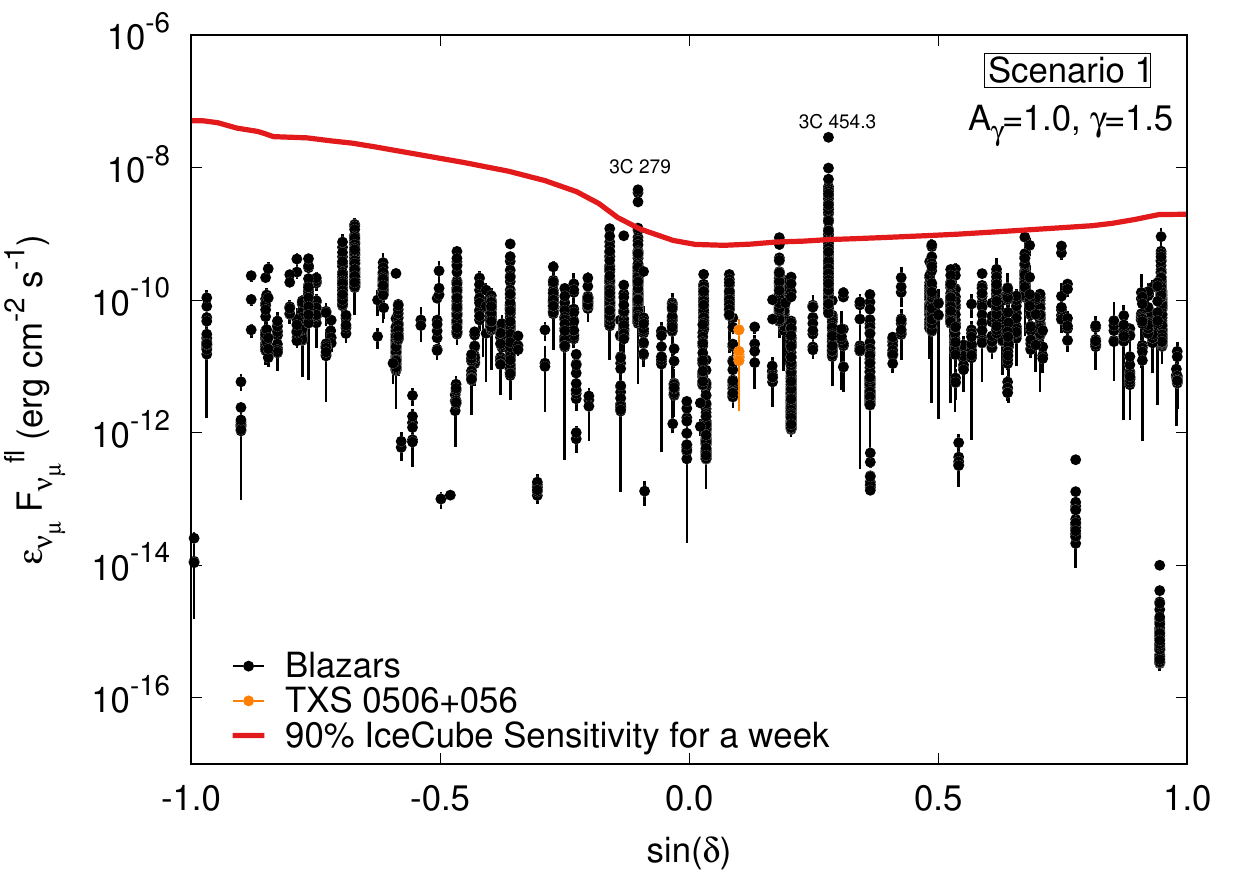}{0.48\textwidth}{(a)}
          \fig{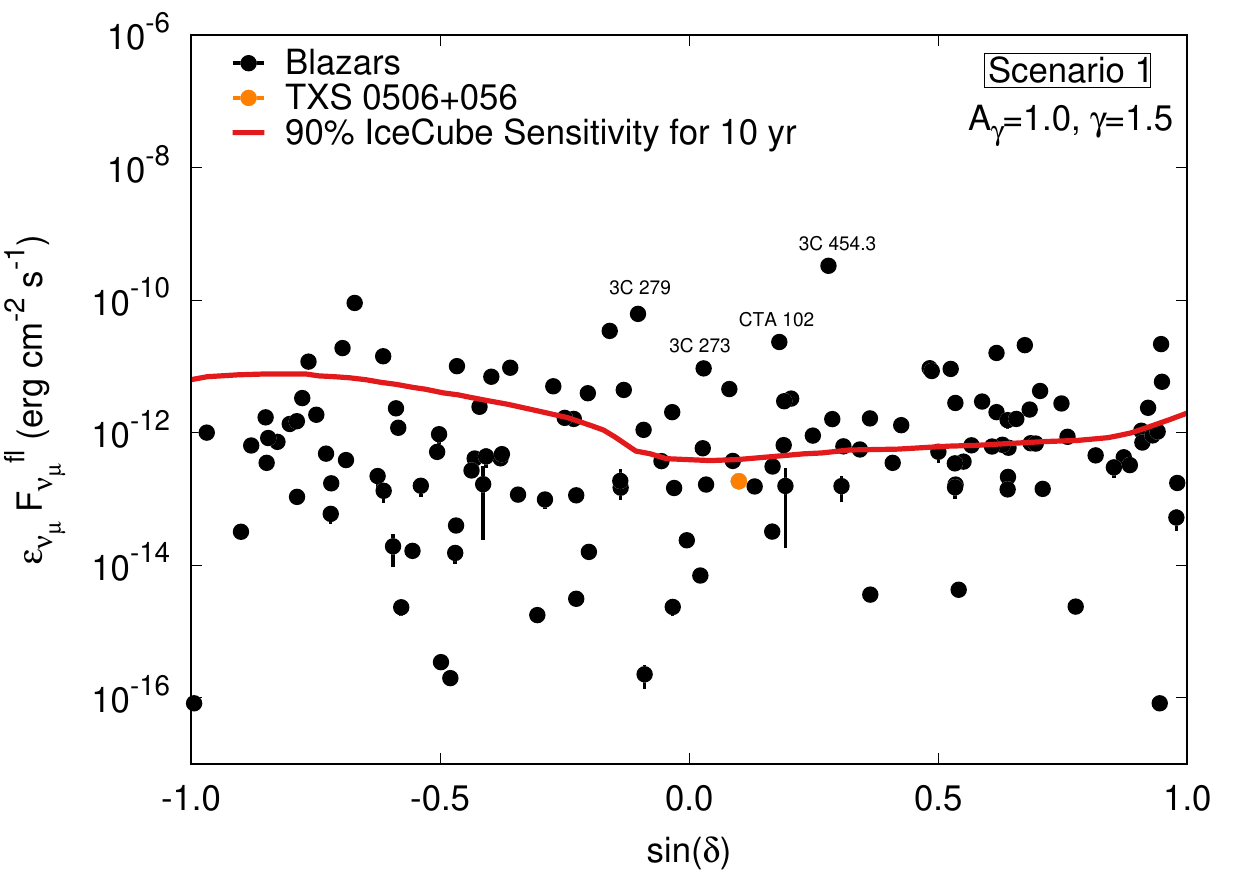}{0.48\textwidth}{(b)}
}
\caption{Estimated muon neutrino flare fluxes of 136 blazars (102 FSRQs, 26 BL Lacs, and 8 BCUs) for a one-week bin (a) and a 10-year bin (b) of scenario 1 with $A_{\gamma} = 1.0$ and ${\gamma} = 1.5$ as a function of ${\sin}({\delta})$ where ${\delta}$ is the source declination. The IceCube 90~\% sensitivities (red solid lines) for a point source with an $E_{\nu}^{-2}$ neutrino spectrum for one week and 10 years are taken from \cite{2019ICRC...36.1026V} and \cite{2020PhRvL.124e1103A}, respectively.}
\label{fig:sindecl2nufl_gqb_A1g15}   
\end{figure*}

\begin{figure*}
\gridline{\fig{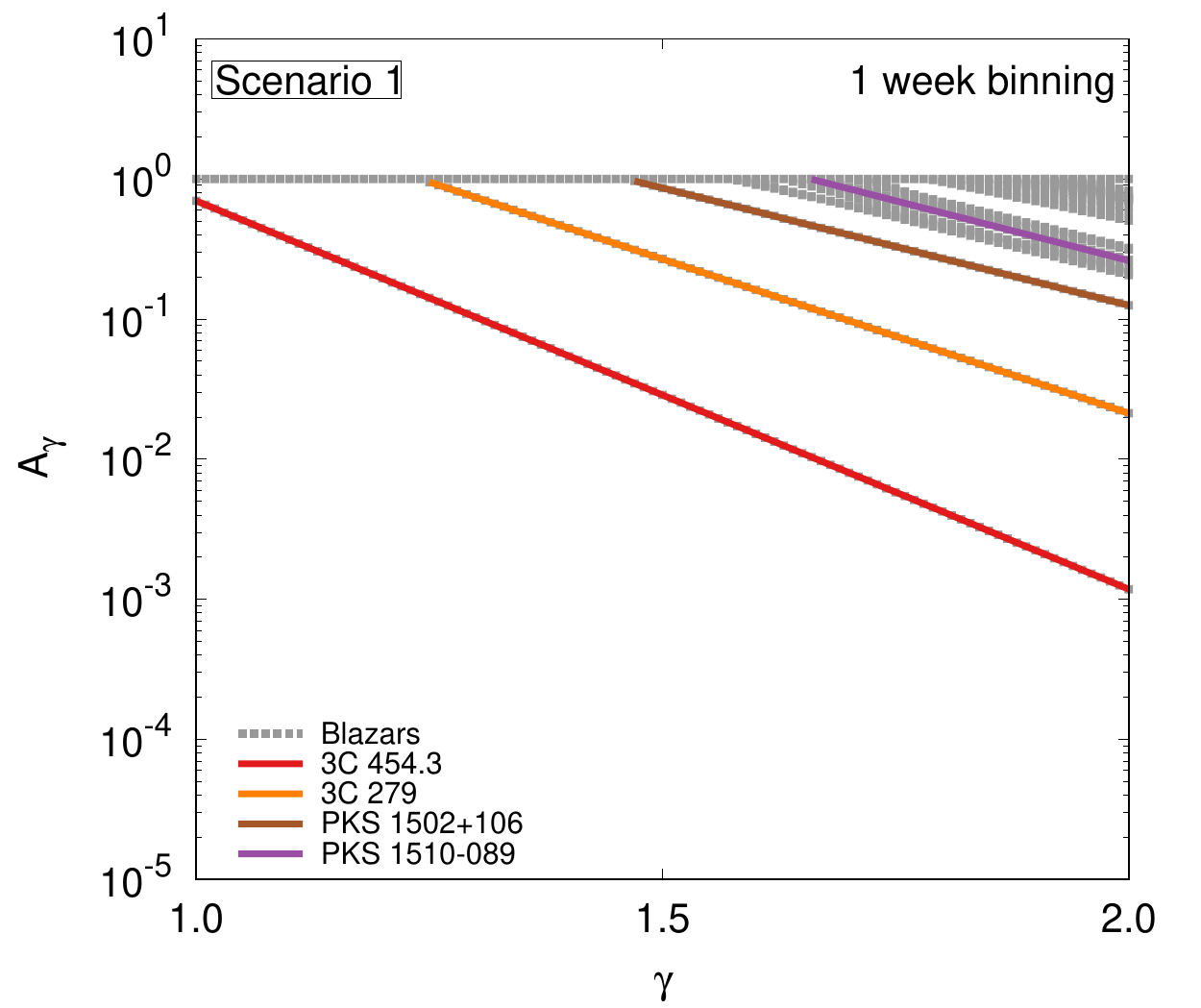}{0.45\textwidth}{}
          \fig{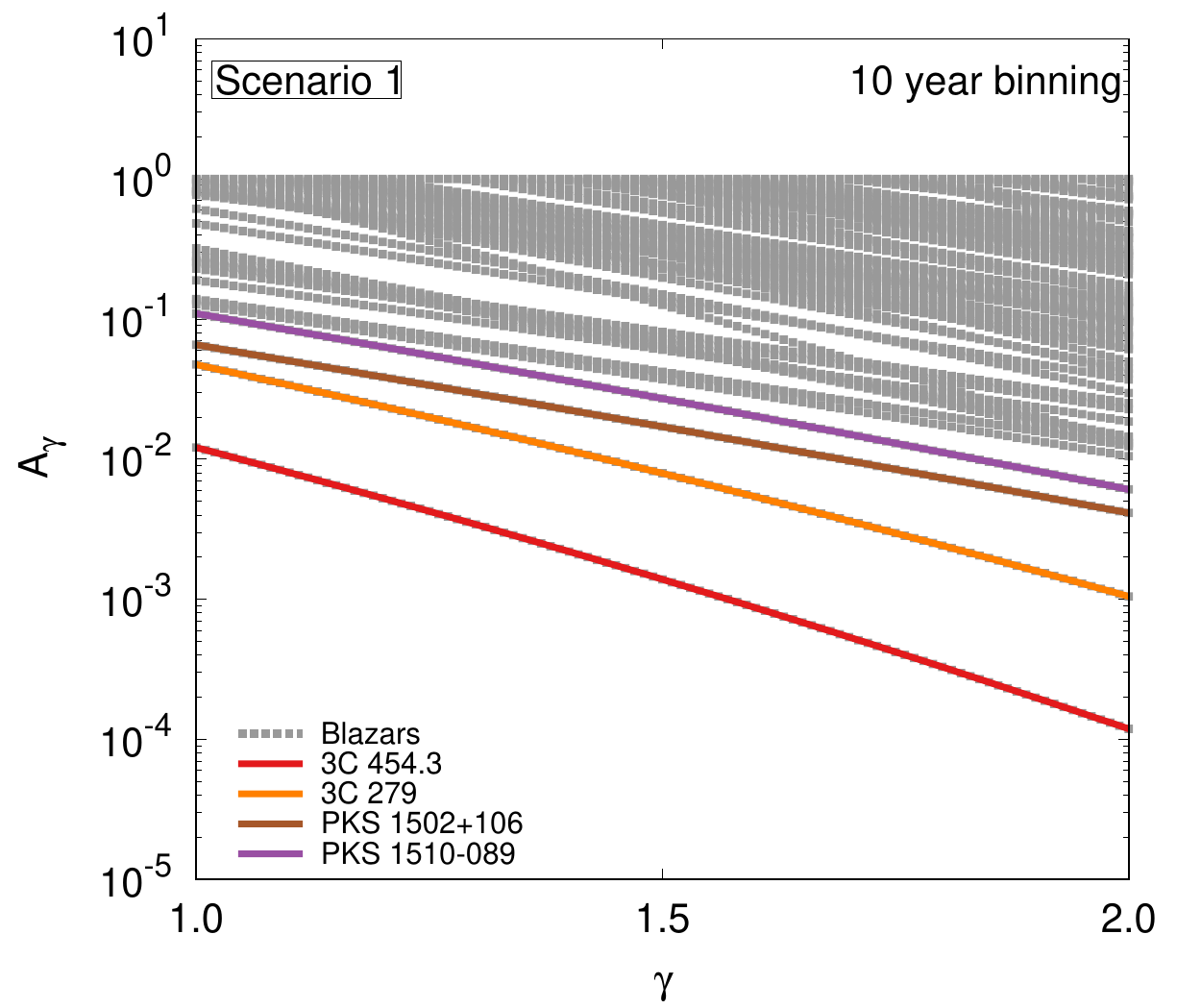}{0.45\textwidth}{}
}
\caption{Upper limits on the normalization parameter $A_{\gamma}$ of the analyzed blazars in gray as a function of the indices $\gamma$ based on the non-detection of neutrino events, compared to the IceCube 90 \% sensitivity for scenario 1. 
The left and right panels show the normalization parameters with one-week binning and 10-year binning, respectively. When sources whose flares are all less than the sensitivity flux, the normalization parameters of $A_{\gamma}$ are set to 1.0. The normalization parameter of the source is adjusted so that the maximum neutrino flux of the source is equal to the sensitivity flux if the flux of at least one flare exceeds the sensitivity.
Several sources with the lowest normalization parameters are plotted with red, blue, brown, and purple lines.}
\label{fig:g2Ag}   
\end{figure*}

\subsection{X-ray proxy (scenario 2)}
This case is motivated by modeling studies of the \txs multi-messenger observations of \txs~\citep[e.g.,][]{Keivani:2018rnh, 2020ApJ...891..115P}, which demonstrated that the neutrino fluxes are bound by the X-ray data. The quiescent muon neutrino flux is benchmarked with the quiescent X-ray flux, $E_X F^{q}_{E_X}$, and the muon neutrino flare flux is written as
\eqb    
    E_{{\nu}_{\mu}}F^{\rm fl}_{{\nu}_{\mu}} = E_{{\nu}_{\mu}}F^{q}_{E_{{\nu}_{\mu}}}  \left( \frac{F^{\rm fl}_{\gamma}}{F^{q}_{\gamma}} \right)^\gamma= A_{X} \, \frac{E_X F^{q}_{E_X}}{3}\left( \frac{F^{\rm fl}_{\gamma}}{F^{q}_{\gamma}} \right)^\gamma
    \label{eq:scenarioX}
\eqe 
where $A_{X}$ is a normalization parameter. For $\gamma = 1.0-2.0$, we can set an upper limit on $A_X$ so that our predictions are consistent with the non-detection of neutrinos from flares of individual blazars. 

For the X-ray data  we use the Open Universe for Blazars, which provides  blazar X-ray light curves based on 14 years of \swift-XRT data \cite{Giommi:2019}. 
The 0.5~keV, 1.0~keV, 1.5~keV, 3.0~keV, and 4.5~keV quiescent fluxes are estimated in the same way as the gamma-ray quiescent level (see Section \ref{sec:quiescent}).
The X-ray quiescent flux, $E_X F^{q}_{E_X}$, is then defined as the average of the quiescent fluxes in these five energies. 
Figure~\ref{fig:xray_quiescent} in Appendix~\ref{sec:xray_quiescent_flux} shows examples of the 1.0~keV X-ray light curves with the respective Bayesian blocks representations and quiescent fluxes. 
By using equation (\ref{eq:scenarioX}), we derive the neutrino flare fluxes of 129 blazars (97 FSRQs, 26 BL Lacs, 6 BCUs), ignoring 16 sources without flaring states found at the $6\sigma$ threshold or without well-defined X-ray quiescent state.

\begin{figure*}
\gridline{\fig{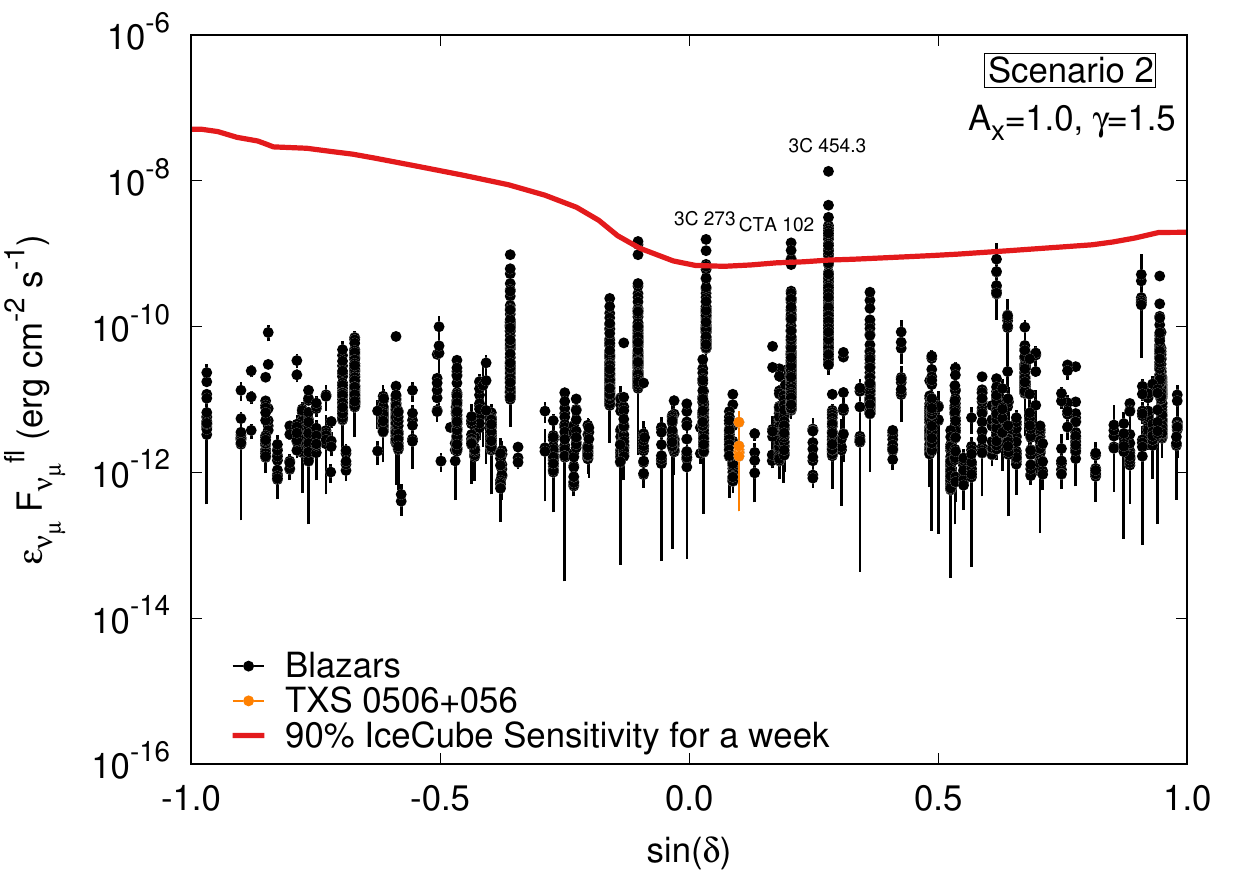}{0.48\textwidth}{(a)}
          \fig{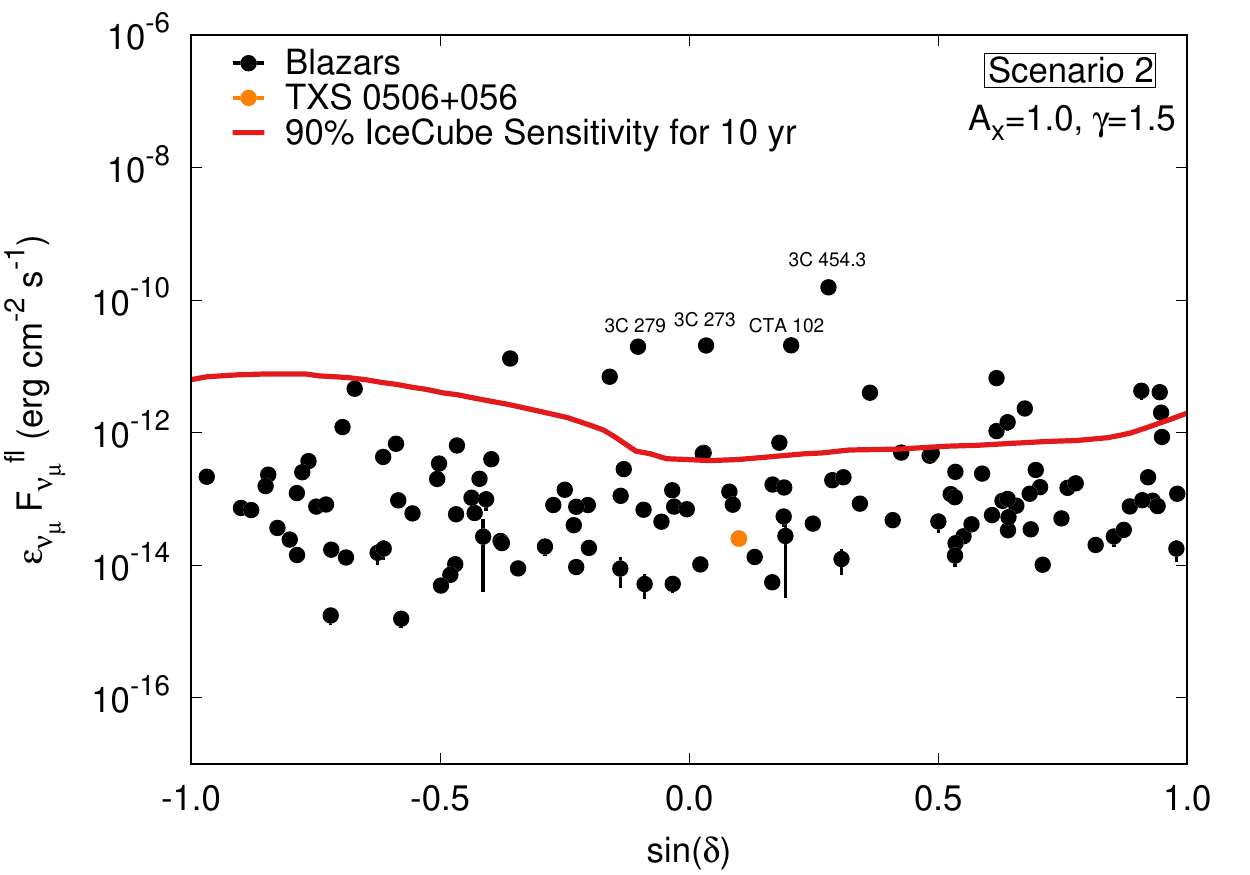}{0.48\textwidth}{(b)}
}
\caption{Estimated muon neutrino flare fluxes of 129 blazars (97 FSRQs, 26 BL Lacs, and 6 BCUs) for a one-week bin (a) and a 10-year bin (b) of scenario 2 with $A_{X} = 1.0$ and ${\gamma} = 1.5$ as a function of the ${\sin}({\delta})$ with declination ${\delta}$. The 90~\%  IceCube sensitivities are the same as in Figure~\ref{fig:sindecl2nufl_gqb_A1g15}.}
\label{fig:sindecl2nuflare_A1g15}   
\end{figure*}

\begin{figure*}
\gridline{\fig{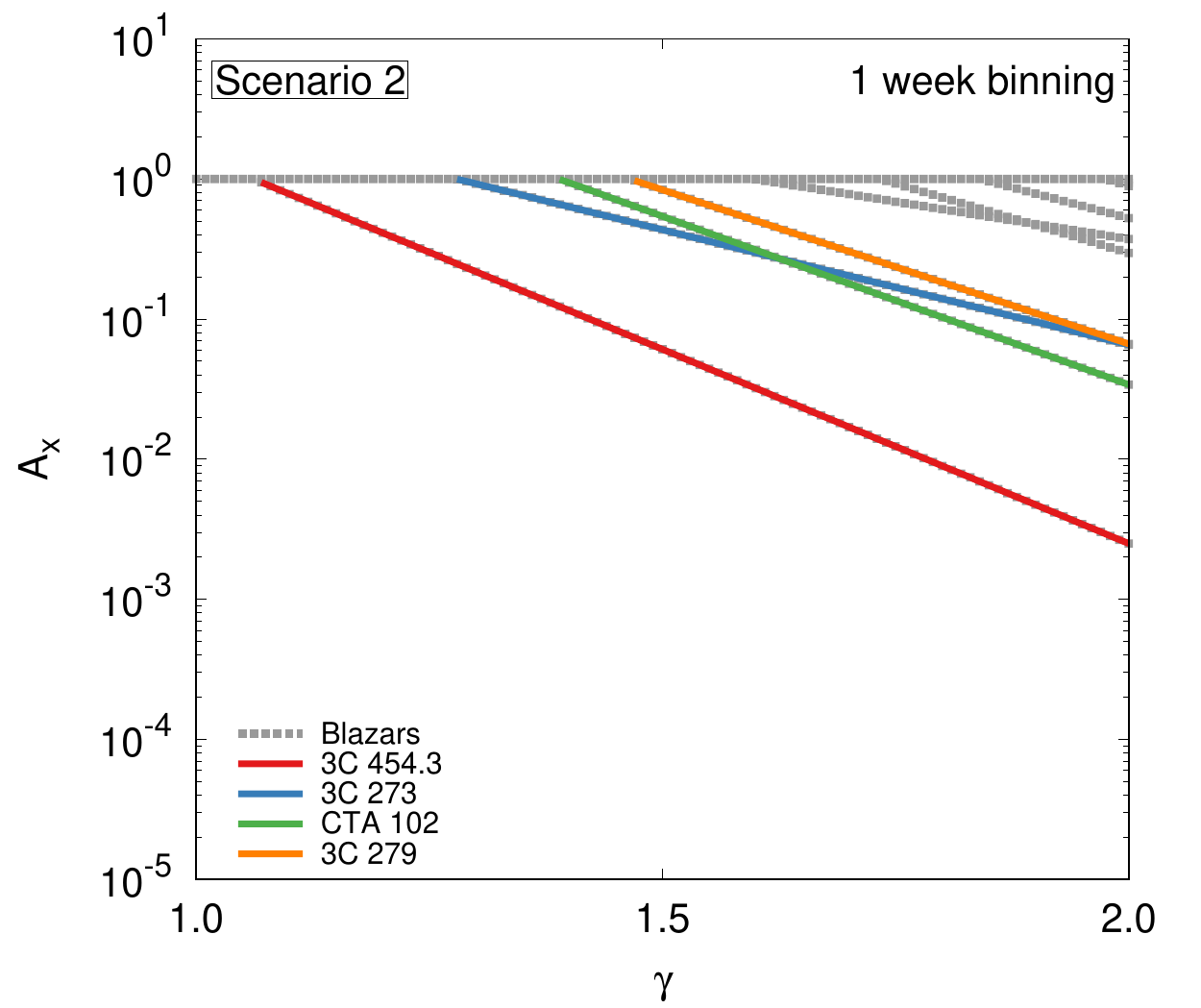}{0.45\textwidth}{}
          \fig{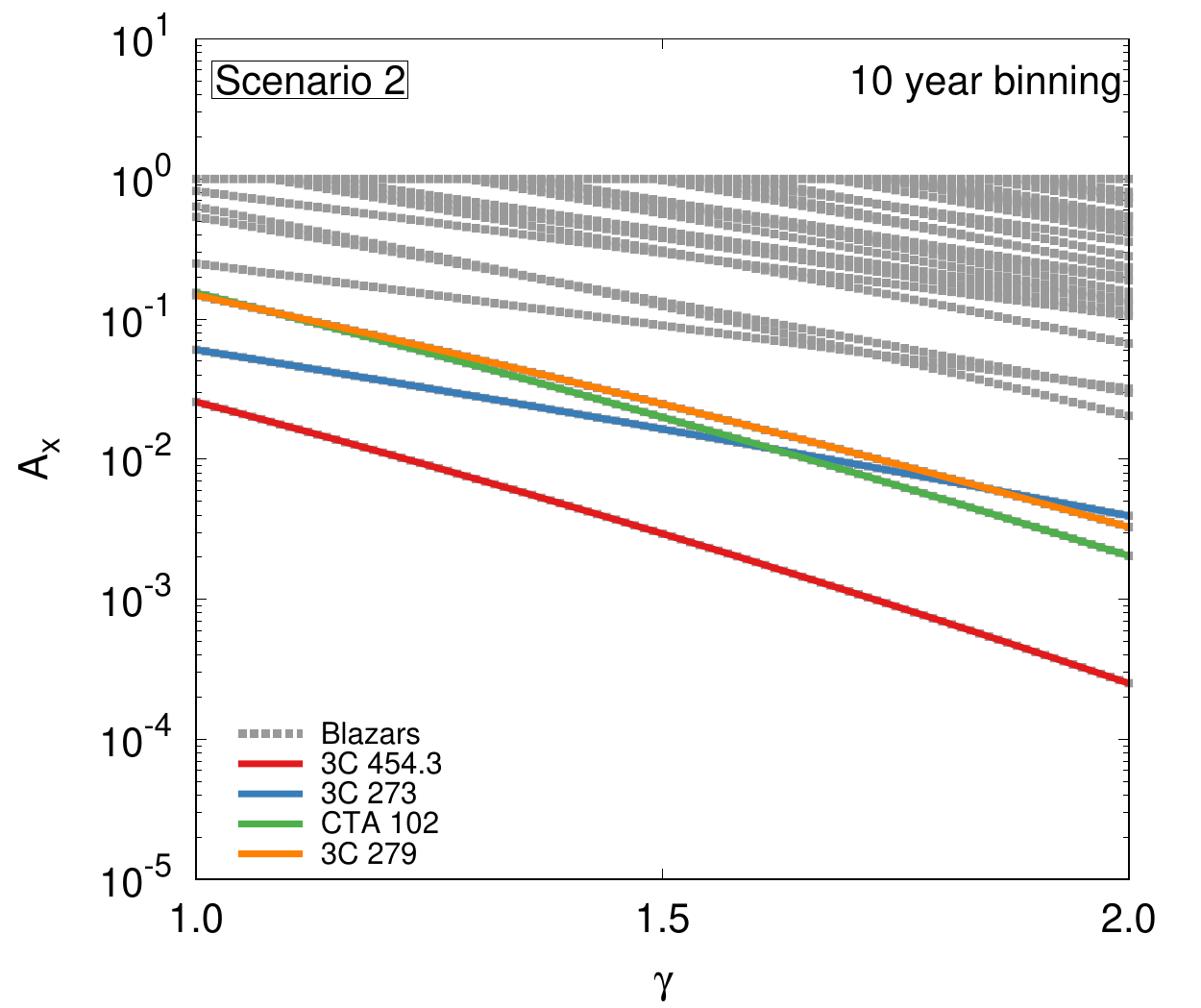}{0.45\textwidth}{}
}
\caption{Upper limits on the normalization parameter $A_{X}$ of the analyzed blazars in gray as a function of the indices $\gamma$ based on the non-detection of neutrino events, compared to the IceCube 90\% sensitivity for scenario 2. Left and right panels show the normalization parameters with one week binning and 10 year binning, respectively. 
For sources whose flares are all less than the sensitivity flux, the normalization parameter $A_{X}$ is set to 1.0. The normalization parameter of the source is adjusted so that the maximum neutrino flux of the source is equal to the sensitivity flux if the flux of at least one flare exceeds the sensitivity.
Several sources with the lowest normalization parameters are plotted with red, blue, green, and orange lines.}
\label{fig:g2Ax}   
\end{figure*}

In Figure~\ref{fig:sindecl2nuflare_A1g15} we plot the estimated muon neutrino flare fluxes from scenario 2 with $A_{X} = 1.0$ and ${\gamma} = 1.5$ as a function of ${\sin}({\delta})$, for two binning choices (panels a and b).
The estimated neutrino flare fluxes in scenario 2 are generally lower by one or two orders of magnitude than those in scenario 1 (compare to Figure~\ref{fig:sindecl2nufl_gqb_A1g15}).
This result is related to the different spectral energy distributions (SEDs) of blazars in our sample. The majority of them are FSRQs, which are typically low-synchrotron peaked sources and their SED has a trough in the X-ray band covered by \swift-XRT. As a result, the X-ray fluxes used as a proxy to scale the neutrino luminosity are lower than the gamma-ray fluxes used in scenario 1 (see also Figure~\ref{fig:gq2xq_scatt}).
This does not apply, however, to MeV blazars and extreme synchrotron peaked sources \citep{2020MNRAS.491.2771C}.
Figure~\ref{fig:g2Ax} shows the normalization parameter $A_{X}$ in gray as a function of the indices $\gamma$ based on the non-detection of neutrino events, compared to the IceCube 90~\% sensitivity for one-week binning (left) and 10-year binning (right). The red, blue, green, and orange lines show sources with the lowest normalization parameters (see inset legend).

\subsection{Impacts of realistic spectra and upper limits}
We also investigate the sensitivity of our results to a non-power-law shape of the neutrino spectrum. To this end, we assume a more realistic, albeit model-dependent neutrino spectrum motivated by leptohadronic models of TXS 0506+056. We compare the expected number of neutrinos to the one expected under the assumption of an $E_\nu^{-2}$ neutrino spectrum, which we have used in the previous sections. We choose the LMBB2b model of~\cite{Keivani:2018rnh} as an illustrative example of a characteristic neutrino spectrum that might result from photomeson production processes in the jet of a blazar during a gamma-ray flare. 

We use equations~(\ref{eq:scenarioG}) and~(\ref{eq:scenarioX}) with a 10-year binning to test this variant of scenarios 1 and 2 respectively, and to place upper limits on the normalization parameters, $A_{\gamma}$ and $A_{X}$. To calculate the expected number of neutrinos, we assume the IC86 point source effective area given in~\cite{IceCube:2016tpw}. 

Columns 4-5 and 6-7 of Table~\ref{tab:scenario3} can also be used to see the effect of the more realistic spectral shape. For TXS~0506+056 and 3C~454.3, the expected number of neutrinos with the more realistic model is a factor of two lower than that of an $E_\nu^{-2}$ neutrino spectrum. Therefore the respective upper limits of the more realistic model are a factor of two less stringent constraints. We thus conclude that the results obtained with our standard assumption of an $E_\nu^{-2}$ spectral shape may be up to a factor of a few more strict than with a different spectral shape for a given total neutrino energy flux. Regardless of the neutrino spectral template, the upper limits on $A^{\rm up}_{\gamma/X}$ are several orders of magnitude more constraining for 3C~454.3 than for TXS~0506+056.

    \begin{table*}
    \caption{Neutrino counts using the LMBB2b model of~\cite{Keivani:2018rnh} (columns 4, 5) and an $E_\nu^{-2}$ neutrino spectrum (columns 6, 7). 
    The assumed upper limit is 2.44 neutrinos in the livetime of IceCube.
    The IC86 point source effective area has been assumed.}
    \centering
    \begin{tabular}{c c c c c c c}
    \hline 
    \hline
    & & & \multicolumn{2}{c}{LMBB2b} & \multicolumn{2}{c}{$E_\nu^{-2}$} \\
    Source & Scenario & $\gamma$ & $N_{\nu_{\mu}+\bar{\nu}_{\mu}}(A_{\gamma/X} = 1$) & $A_{\gamma/X}^{\rm up}$ (90\% CL) &  $N_{\nu_{\mu}+\bar{\nu}_{\mu}}(A_{\gamma/X} = 1$ ) & $A_{\gamma/X}^{\rm up}$ (90\% CL)  \\
    \hline
    TXS 0506+056 & 1 & 1.0 & 2 & 1 & 4 & 1 \\
    TXS 0506+056 & 1 & 1.5 & 6 & 1 & 10 & 1 \\
    TXS 0506+056 & 1 & 2.0 & 10 & 1 & 20 & 1 \\
    TXS 0506+056 & 2 & 1.0 & 0.3 & 1 & 0.5 & 1\\
    TXS 0506+056 & 2 & 1.5 & 0.9 & 1 & 1.0 & 1\\
    TXS 0506+056 & 2 & 2.0 & 2 & 1 & 3 & 1\\
    3C 454.3 & 1 & 1.0 & 20 & 0.1 & 70 & 0.04 \\
    3C 454.3 & 1 & 1.5 & 200 & $10^{-2}$& 600 & $4 \times 10^{-3} $\\
    3C 454.3 & 1 & 2.0 & 2000 & $10^{-3}$ & 6000 & $4 \times 10^{-4} $\\
    3C 454.3 & 2 & 1.0 & 10 & 0.2 & 30 & 0.08 \\
    3C 454.3 & 2 & 1.5 & 100 & $2\times10^{-2}$& 300 & $8 \times 10^{-3} $\\
    3C 454.3 & 2 & 2.0 & 1000 & $2\times10^{-3}$& 3000 & $8 \times 10^{-4} $ \\
        \hline
    \end{tabular}
    \label{tab:scenario3}
\end{table*}

\subsection{Contribution of blazar flares to the all-sky neutrino flux}
\label{sec:contribution2INB}

There are two ways to examine the contribution of blazars to the all-sky neutrino flux. One is the stacking method considering the directions of IceCube neutrino events \citep{2017ApJ...835...45A,Hooper19,IceCube:2022zbd}, and the other is the clustering analysis \citep{Murase:2016}.  
Instead, our flux stacking method uses the estimated muon neutrino fluxes, not directly using the spatial correlation between the directions of IceCube neutrino events and blazars, but using the IceCube sensitivity at the declination of each blazar in our sample.

As shown in Figures~\ref{fig:sindecl2nufl_gqb_A1g15} and \ref{fig:sindecl2nuflare_A1g15}, some neutrino flare fluxes with $A_{X} = 1.0$ and $A_{\gamma} = 1.0$ exceed the IceCube 90~\% sensitivity. The normalization parameters of $A_{\gamma}$ and $A_{X}$ are set to 1.0 for the sources whose flares are \textit{all} less than the sensitivity flux. However, the normalization parameter of the source is reduced, if the flux of at least one flare exceeds the sensitivity,  in a way that the maximum neutrino flux of the source is equal to the sensitivity flux.

In this work, we estimate the upper limits on the diffuse isotropic neutrino flux through flux stacking analyses of blazar flares. Our sample consists of bright gamma-ray blazars from the 4LAC catalogue as shown in Figure~\ref{fig:z2F_scatthist}. 
There are fainter blazars in the 4LAC catalog, and there ought to be a lot of not-yet-detected blazars in the Universe. 

To obtain the upper limits on the all-sky neutrino flux, the sum of the estimated neutrino fluxes from the flaring blazars in our sample is divided by $4{\pi}$ and then multiplied by two factors for the completeness correction. 
First, the correction factor of all 4LAC blazars and the blazars of our sub-sample is obtained by the fraction of the total gamma-ray energy fluxes in 0.1 -- 316~GeV (raised to the power of ${\gamma}$) of 4LAC blazars to the sample blazars: 
\begin{equation}
{ \sum_{i=1}^{N_c}( {\rm Energy Flux}_i)^{\gamma} } / { \sum_{j=1}^{N_s} ({\rm Energy Flux}_j)^{\gamma} }
\label{eq:sample2lac}
\end{equation}
where $N_c$ is the number of blazars in the 4LAC catalog, $N_s$ is the number of blazars in our sample, ${\rm Energy Flux}_{i,j}$ are time-average gamma-ray energy fluxes in 0.1 -- 316~GeV.  
The second correction factor for the blazar population,  which includes too distant or too low in gamma-ray luminosity sources, and the 4LAC blazars, is obtained by completeness correction factors with $\gamma$, which are taken from Figure 1 of \cite{Yuan2020}. Since the analysis of \cite{Yuan2020} originates from the 3LAC catalog, our results are slightly conservative when applying it to the 4LAC catalog. The correction factors are presented in Table~\ref{tab:correction_factor}.

\begin{table}
\begin{center}
    \caption{Correction factors for determining the contribution of blazar flares to the all-sky neutrino flux.}
    \label{tab:correction_factor}
    \begin{tabular}{c | c c c}
    \hline \hline
             & \multicolumn{2}{c}{4LAC/Sample \footnote{The correction factor of all the 4LAC FSRQs/BL Lacs}}
             & Total/4LAC \footnote{The correction factor of the whole population of blazars} \\
    ~~$\gamma$~~ &  ~~FSRQs~~  & ~~BL Lacs~~  &   \\
    \hline 
    1.0     &  1.76  & 4.24  & 3.87 \\
    1.5     &  1.22  & 2.21  & 1.09 \\
    2.0     &  1.06  & 1.60  & 1.01 \\
    \hline
    \end{tabular}
\end{center}
\end{table}

\begin{figure*}
\gridline{\fig{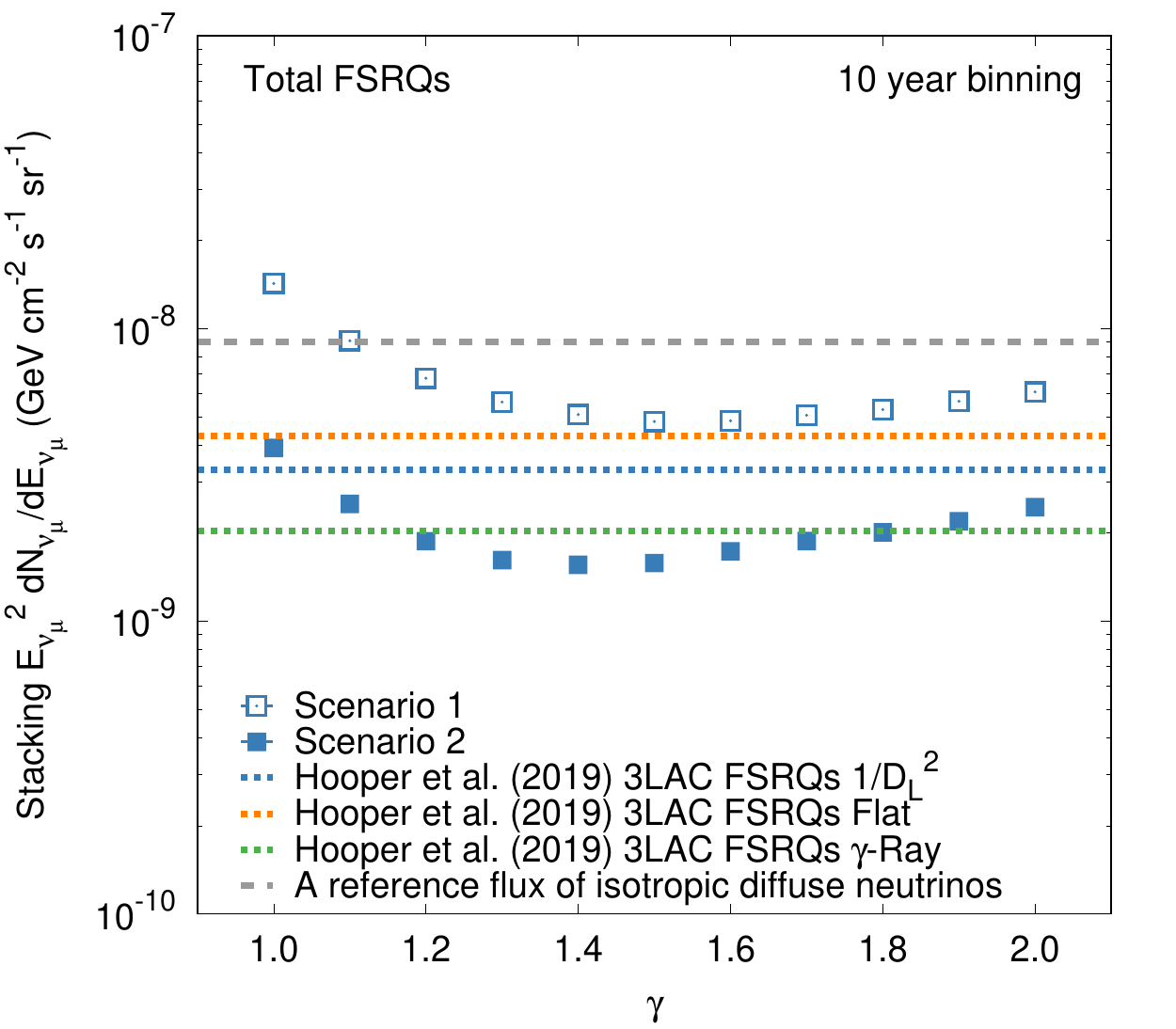}{0.45\textwidth}{}
          \fig{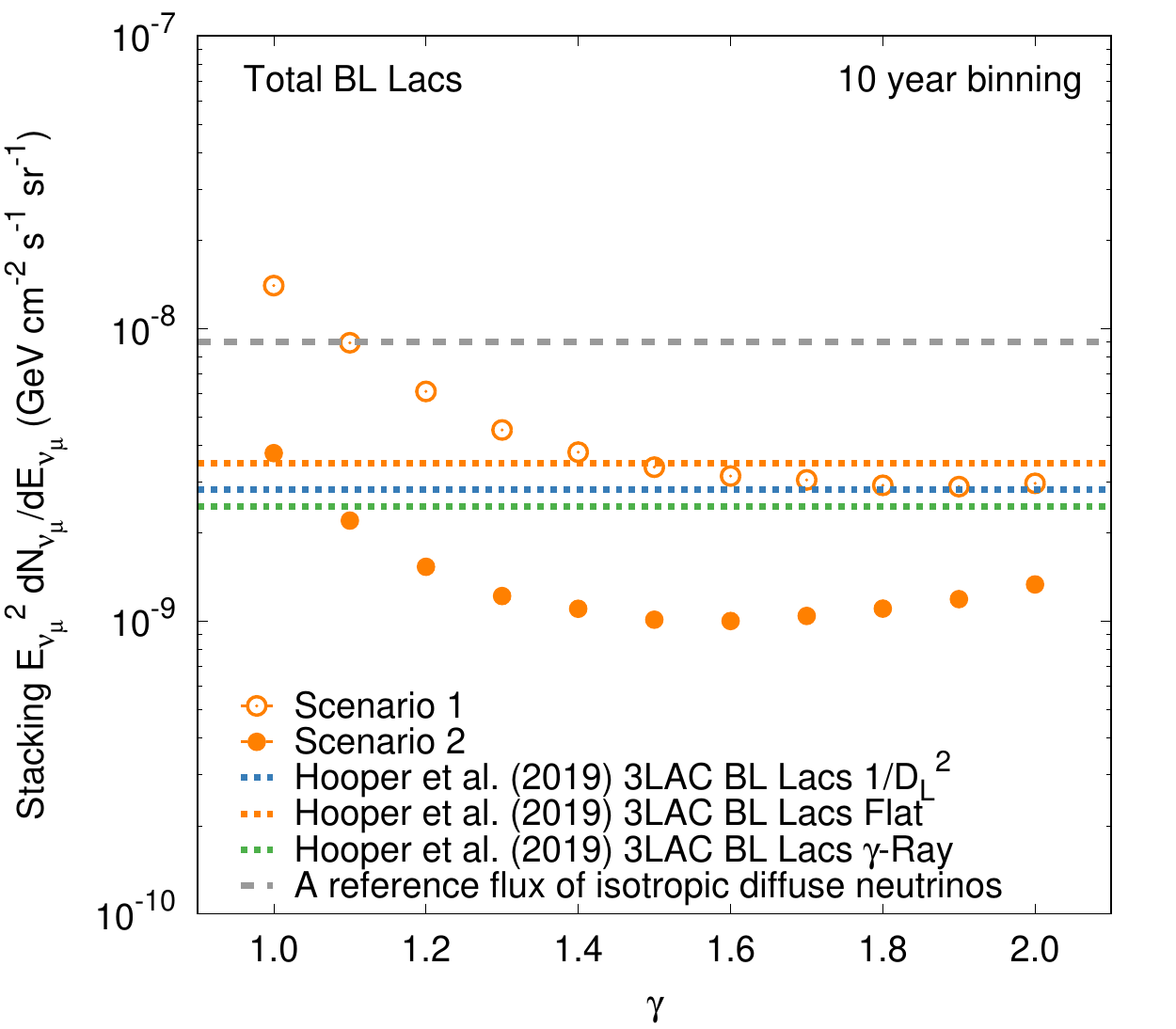}{0.45\textwidth}{}
}
\caption{The total muon neutrino fluxes of gamma-ray flaring FSRQs and BL Lacs as a function of $\gamma$. The total fluxes of muon neutrinos are estimated from our samples, multiplied by correction factors of the contribution of the 4LAC blazars and not-yet-detected blazars that are too gamma-ray faint to be included in the 4LAC catalog. The results of scenarios 1 and 2 are shown in blue solid squares and orange solid circles, compared to a reference flux of all-sky muon neutrinos (gray dashed line)~\citep{2016ApJ...833....3A}. The three upper limits of \cite{Hooper19} derived for different assumptions about the source evolution ($1/D_{L}^{2}$ , Flat, $\gamma$-Ray), and reduced to $1/3$ for muon neutrinos, are overplotted for comparison.}
\label{fig:g2sum_nufl_hooper}
\end{figure*}

Figure ~\ref{fig:g2sum_nufl_hooper} shows the total fluxes of muon neutrinos estimated from our bright gamma-ray FSRQs and BL Lacs, multiplied by correction factors of the contribution of the 4LAC blazars and still unresolved blazars that are too gamma-ray faint to be included in the 4LAC catalog. 
The total muon neutrino fluxes of FSRQs and BL Lacs of scenarios 1 and 2 are plotted as a function of $\gamma$, compared to a reference flux of the isotropic diffuse muon neutrino flux ~\citep{2016ApJ...833....3A}  and the three upper limits derived by \cite{Hooper19}, which are reduced to $1/3$ for muon neutrinos. 
As the neutrino energy fluxes are enhanced by the weighting factor $\gamma$, the sum of the neutrino energy fluxes of the sample blazars gradually increases with $\gamma$. On the other hand, the correction factors sharply decrease with $\gamma$ up to ${\gamma} ~{\simeq}~ 1.5$ and stay nearly constant for $\gamma ~{\gtrsim}~ 1.5$ as shown in Table~\ref{tab:correction_factor}. Therefore, the total neutrino energy fluxes, namely the product of the sum of the neutrino energy fluxes of the sample blazars and the correction factors, decrease with $\gamma$ up to ${\gamma} ~{\simeq}~ 1.5$ and turn to increase gradually with $\gamma$ for $\gamma ~{\gtrsim}~ 1.5$ as shown in scenario 1 of Fig.~\ref{fig:g2sum_nufl_hooper}. On the other hand, the total neutrino energy fluxes are maintained nearly constant for $\gamma ~{\gtrsim}~ 1.5$ as shown in scenario 2 of Fig.~\ref{fig:g2sum_nufl_hooper}, since the gamma-ray normalization factors $A_{\gamma}$ are strongly suppressed in tens of blazars shown in Fig.~\ref{fig:g2Ag} (right). 
The upper limits of scenario 1 are in agreement with those of \cite{Hooper19} for $\gamma \gtrsim 1.5$ even though the analyses and assumptions are different. We predict weaker upper limits though for smaller values of $\gamma$. This is because the second correction factor should be larger for small values of $\gamma$ \citep{Yuan2020}. We also note that the upper limits of scenario 2 are tighter than those of scenario 1 for all values of $\gamma$.

\begin{figure}
\begin{center}
\includegraphics[width=0.45\textwidth]{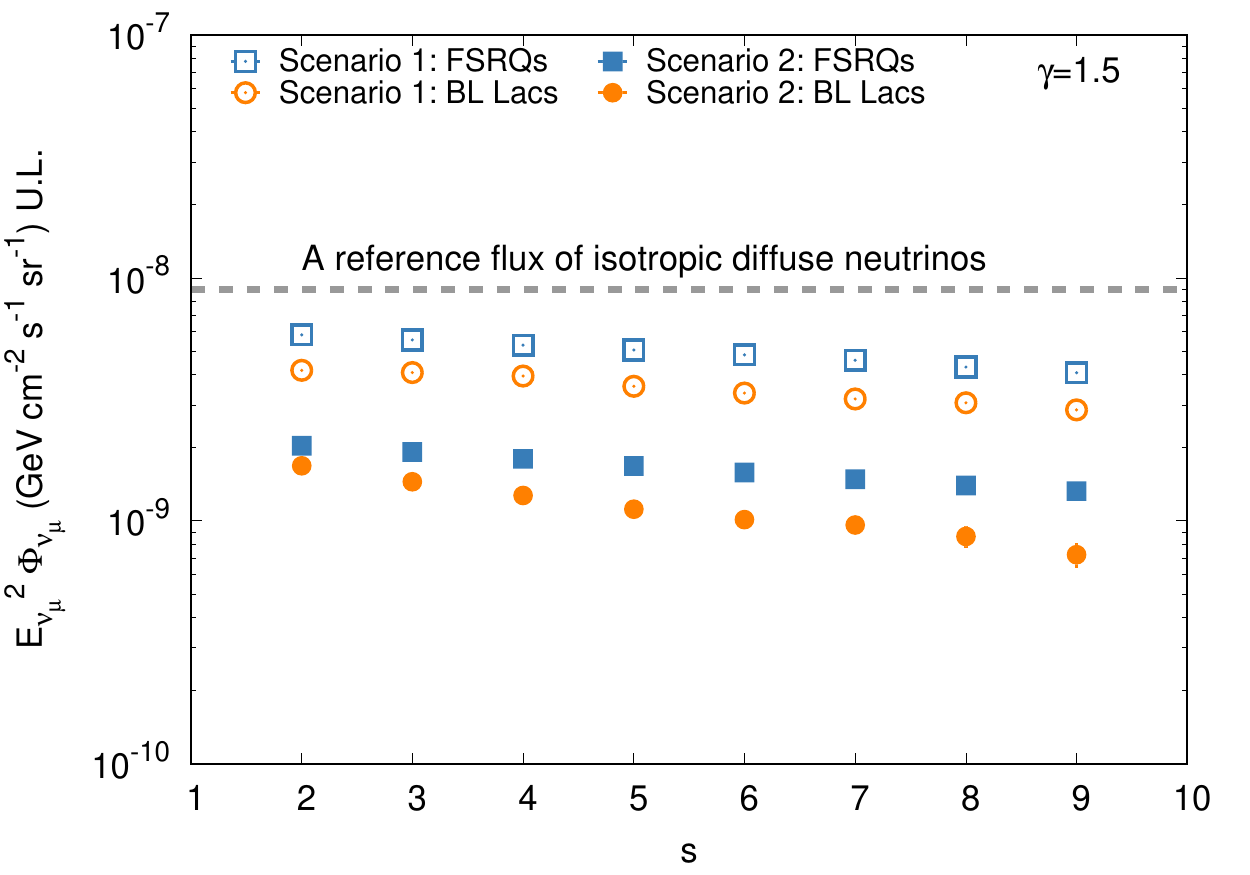}
\end{center}
\caption{The upper limits of $E_{{\nu}_{\mu}}^{2} {\Phi}_{{\nu}_{\mu}}$ of a 10-year binning as a function of the flare significance $s$ above the quiescent level for FSRQs (open blue squares) and BL Lacs (open orange circles) of scenario 1  and FSRQs (solid blue squares) and BL Lacs (solid orange circles) of scenario 2 with $\gamma = 1.5$. The dashed horizontal line shows a reference flux of the isotropic diffuse muon neutrinos~\citep{2016ApJ...833....3A}.}
\label{fig:sig2inb_sum}
\end{figure}

So far, we have defined gamma-ray flares with the $6{\sigma}$ level ($s=6$) above the quiescent gamma-ray flux. Figure~\ref{fig:sig2inb_sum} presents the  upper-limit values of $E_{{\nu}_{\mu}}^{2} {\Phi}_{{\nu}_{\mu}}$ as a function of the flare significance $s$ above the quiescent level for FSRQs and BL Lacs in both scenarios with $\gamma = 1.5$. The upper limits of FSRQs and BL Lacs are similar both for scenarios 1 and 2 with each other. The upper limits decrease with the significance levels $s$, becoming $\sim$70~\% from $s = 2$ to $9$ for FSRQs and BL Lacs in scenario 1.

\section{Summary and Conclusions}\label{sec:summary}
We analyzed 145 bright gamma-ray blazars among the 4LAC blazars. Their flare duty cycles and energy fractions represent power-law-like distributions, correlating strongly with each other. We found a significant difference between blazar sub-classes for the flare duty cycles at the 5~\% significant level.
The flare duty cycles and energy fractions also do not correlate with the quiescent gamma-ray luminosities. 
Using monthly-binned light curves of the 2FGL catalog, \cite{2011ApJ...743..171A} showed that bright FSRQs and BL Lacs have flare duty cycles of about $0.05 - 0.10$. Our flare duty cycles of the weekly-averaged time bin light curves show much broader distributions for both blazar sub-classes, ranging from 0.0 to 0.6. 
\cite{2017ApJ...846...34A} present the second catalog of flaring gamma-ray sources (2FAV), whose Table 2 shows the number of weekly-bin flares detected for each source for the first 387 weeks of Fermi observations. According to Table 2 in \cite{2017ApJ...846...34A}, the flare duty cycles of the FAVA analysis appear to suppress less than ${\sim}0.2$, while our flare duty cycles extend to ${\sim}0.6$. As described in appendix \ref{sec:fava}, the photon flux distribution from the FAVA analysis is likely to systematically underestimate the flaring contribution in the total energy output of blazars. For the blazars analized in our sample, we find the power-law indices $\alpha$ of the photon flux distributions are less than 2.5. This suggests that high-energy neutrinos of blazars might be produced mainly during the flare phase.

We estimated muon neutrino flare fluxes by using a scaling law of $L_{\nu}^{\rm fl} = L_{\nu}^{q} (L_{\gamma}^{\rm fl}/L_{\gamma}^{q})^{\gamma}$. The quiescent neutrino flux is benchmarked with two proxies: the quiescent ${\gamma}$-ray flux and the quiescent X-ray flux with the normalization parameters of $A_{\gamma}$ and $A_{X}$, whose upper limits are 1.0. By comparing the estimated muon neutrino flare fluxes to the IceCube 90\% sensitivities, $A_{\gamma}$ and $A_{X}$ are restricted to values much lower than 1.0 for several tens of blazars. 
As shown in Figure~\ref{fig:sindecl2nuflare_A1g15}, the predicted neutrino flaring flux for \txs is below the sensitivity in scenario 2, which is motivated by this source. The expected number of muon neutrinos from \txs is $\sim1$ as presented in Table~\ref{tab:scenario3}, which is below the 90\% sensitivity corresponding to $\sim2.4$ events.

The origin of all-sky neutrinos observed in IceCube is one of the most important puzzles in high-energy neutrino astrophysics. 
We found that scenarios 1 and 2 suggest that no more than ${\sim}50$~\% and ${\sim}14$~\% of the all-sky neutrino flux can originate from gamma-ray flares of FSRQs and BL Lacs, respectively. A more realistic neutrino spectrum than the usual $E_{\nu}^{-2}$ power law yields upper limits that are a factor of two less constraining. The upper limits are consistent with those obtained the previous literature despite different methods and assumptions.


\section*{Acknowledgments}
We thank Chengchao Yuan for providing the data of the completeness factor. 
K.M. is supported by the NSF Grant No.~AST-1908689, No.~AST-2108466 and No.~AST-2108467, and KAKENHI No.~20H01901 and No.~20H05852.  M.P. acknowledges support from the MERAC Fondation through the project THRILL and from the Hellenic Foundation for Research and Innovation (H.F.R.I.) under the ``2nd call for H.F.R.I. Research Projects to support Faculty members and Researchers'' through the grant number 3013 (UNTRAPHOB).

\bibliography{flares}{}
\bibliographystyle{aasjournal}

\appendix
\renewcommand{\thefigure}{\Alph{section}.\arabic{figure}}
\setcounter{figure}{0}
\renewcommand{\thetable}{\Alph{section}.\arabic{table}}
\setcounter{table}{0}

\section{List of Our Sources}

A list of the sources analyzed in this work is given in Table \ref{tab:source_list}, which is available in its entirety in machine-readable form. A partial list is presented here. 

\begin{longtable}{l l c c c c l}
\caption{List of our sources}\\
\hline
4FGL Name & Object Name & RAJ2000 & DEJ2000 & Optical Class & SED Class & $z$ \\
\hline
4FGL J0001.5+2113 &  TXS 2358+209 & 0.3815 & 21.2183 & FSRQ & ISP & 1.106 \\
4FGL J0017.5-0514 &  PMN J0017-0512 & 4.3949 & -5.2347 & FSRQ & LSP & 0.227 \\
4FGL J0038.2-2459 &  PKS 0035-252 & 9.5652 & -24.9899 & FSRQ & LSP & 1.196 \\
4FGL J0102.8+5824 &  TXS 0059+581 & 15.701 & 58.4092 & FSRQ & LSP & 0.644 \\
4FGL J0108.6+0134 &  4C +01.02 & 17.1695 & 1.5819 & FSRQ & LSP & 2.099 \\
4FGL J0112.8+3208 &  4C +31.03 & 18.2227 & 32.1399 & FSRQ & LSP & 0.603 \\
4FGL J0132.7-1654 &  PKS 0130-17 & 23.176 & -16.9103 & FSRQ & LSP & 1.02 \\
4FGL J0133.1-5201 &  PKS 0131-522 & 23.2938 & -52.0202 & FSRQ & LSP & 0.925 \\
4FGL J0210.7-5101 &  PKS 0208-512 & 32.6946 & -51.0218 & FSRQ & LSP & 1.003 \\
$\cdots$          &  $\cdots$     & $\cdots$ & $\cdots$  & $\cdots$  & $\cdots$  & $\cdots$  \\
\hline
\multicolumn{7}{p{0.85\textwidth}}{Notes. Optical and SED classes are from Fourth LAT AGN Catalog (4LAC) \cite{Fermi4LAC:2020} . Redshifts are from the 4LAC and SIMBAD Astronomical Database \cite{SIMBAD:2019ASPC}.
A portion is shown here. (This table is available in its entirety in machine-readable form.)}
\label{tab:source_list}
\end{longtable}

\section{Examples of 1-316~GeV gamma-ray light curves} \label{sec:lcHE}

\begin{figure*}[h]
\gridline{          \fig{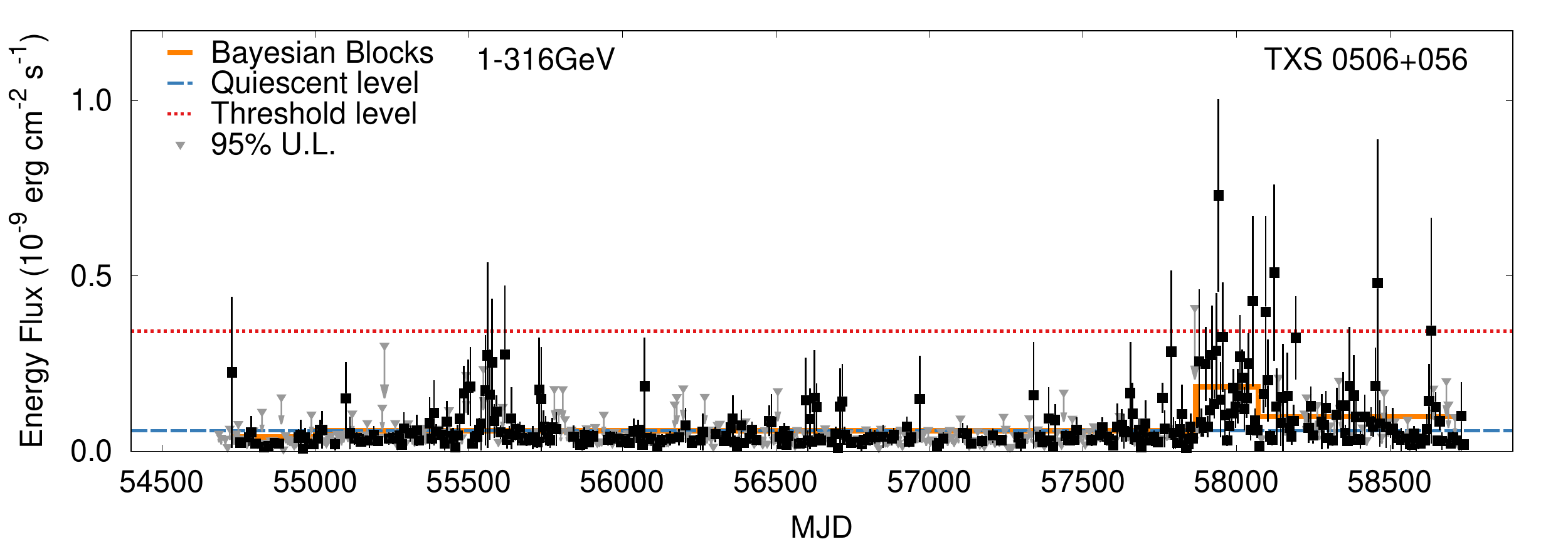}{0.49\textwidth}{(a) TXS 0506+056}
\fig{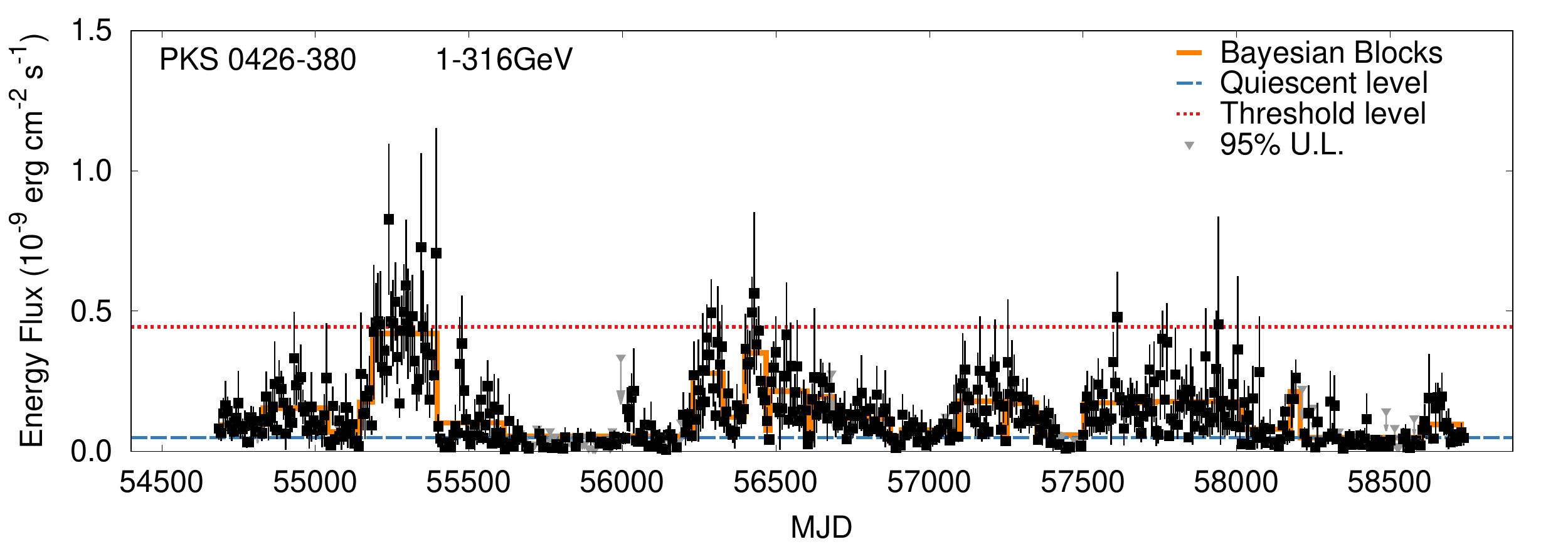}{0.49\textwidth}{(b) PKS 0426-380}
}
\gridline{\fig{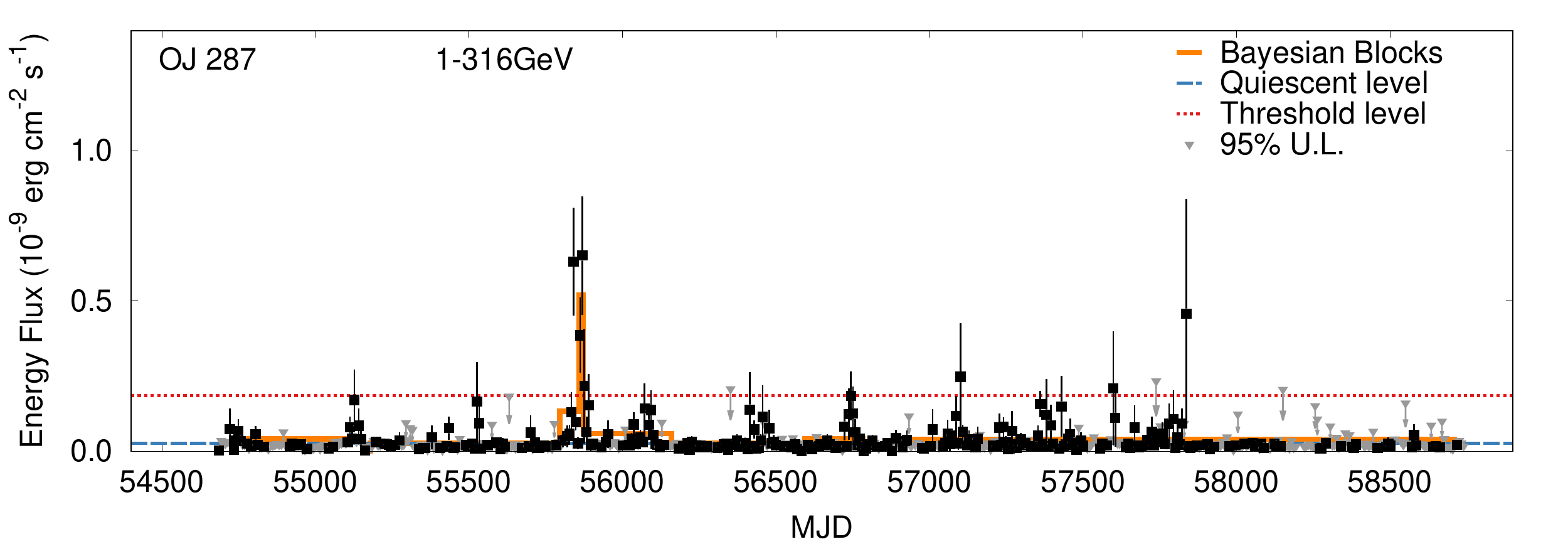}{0.49\textwidth}{(c) OJ 287}
          \fig{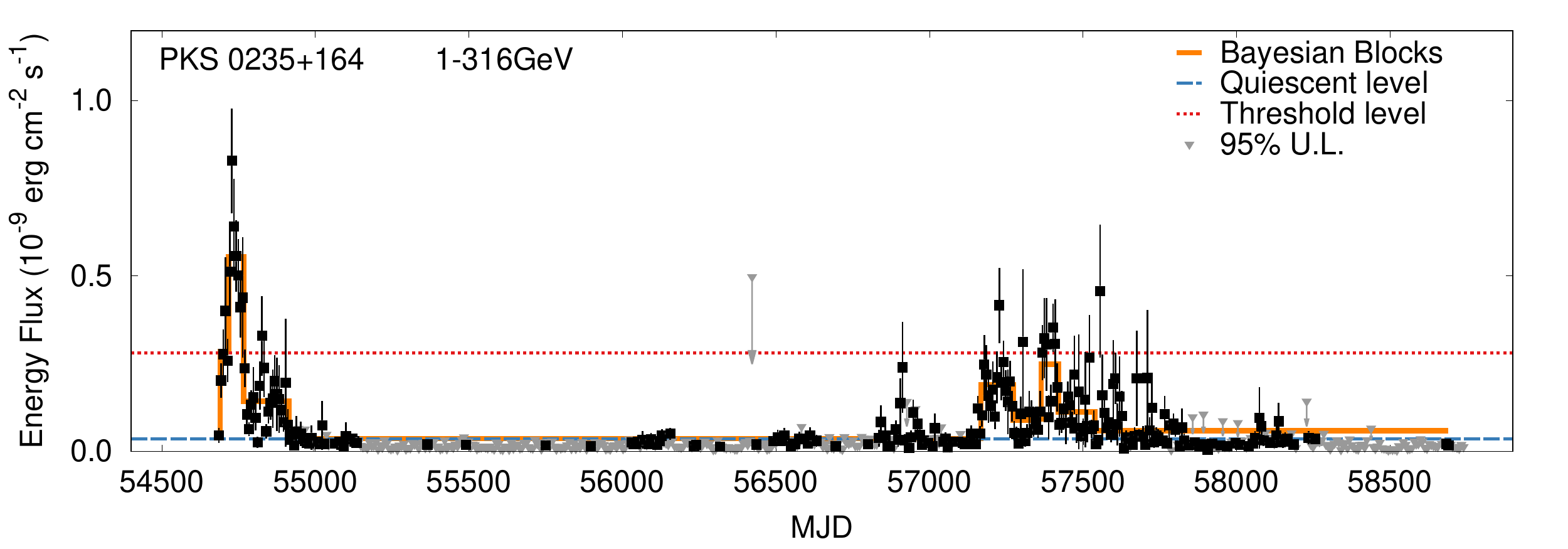}{0.49\textwidth}{(d) PKS 0235+164}
          }
\caption{
Indicative examples of weekly-binned blazar gamma-ray light curves (black points) in the 1-316 GeV energy range. 
Grey triangles indicate 95\% upper limits. See the caption of Figure~\ref{fig:lc} in details. 
\label{fig:lcHE}}
\end{figure*}

Figure \ref{fig:lcHE} presents examples of weekly-binned {\fermi} LAT light curves in the higher energy range of 1-316 GeV (black points). The Bayesian blocks representations and the quiescent flux levels, which are derived by using the method described in section \ref{sec:quiescent}, are also shown in solid orange lines and dashed cyan lines.

\section{Gamma-ray quiescent flux levels with time binning} \label{sec:time_binning}

As mentioned in Section \ref{sec:flaring_rates}, even though there is no widely accepted agreement on how to determine the quiescent flux, it is reasonable to assume that it remains low and constant. When the gamma-ray flux is assumed to be low and constant, the quiescent gamma-ray flux does not increase with coarser time binning unless the flaring state is in effect. However, the coarser time binning should cause an increase in the quiescent gamma-ray flux, accounting for the flaring fluxes. Figure~\ref{fig:binN2qbr} presents the mean gamma-ray quiescent fluxes of 106 FSRQs, 31 BL Lacs, and 8 BCUs with standard deviations as a function of time bins. The quiescent fluxes remain stable below $\sim$20 weeks and gradually increase above $\sim$20 weeks, indicating that they can reasonably be determined to be relatively low and constant using a 1-week time bin.

\setcounter{figure}{0}
\begin{figure*}[h]
\begin{center}
\includegraphics[width=0.45\textwidth]{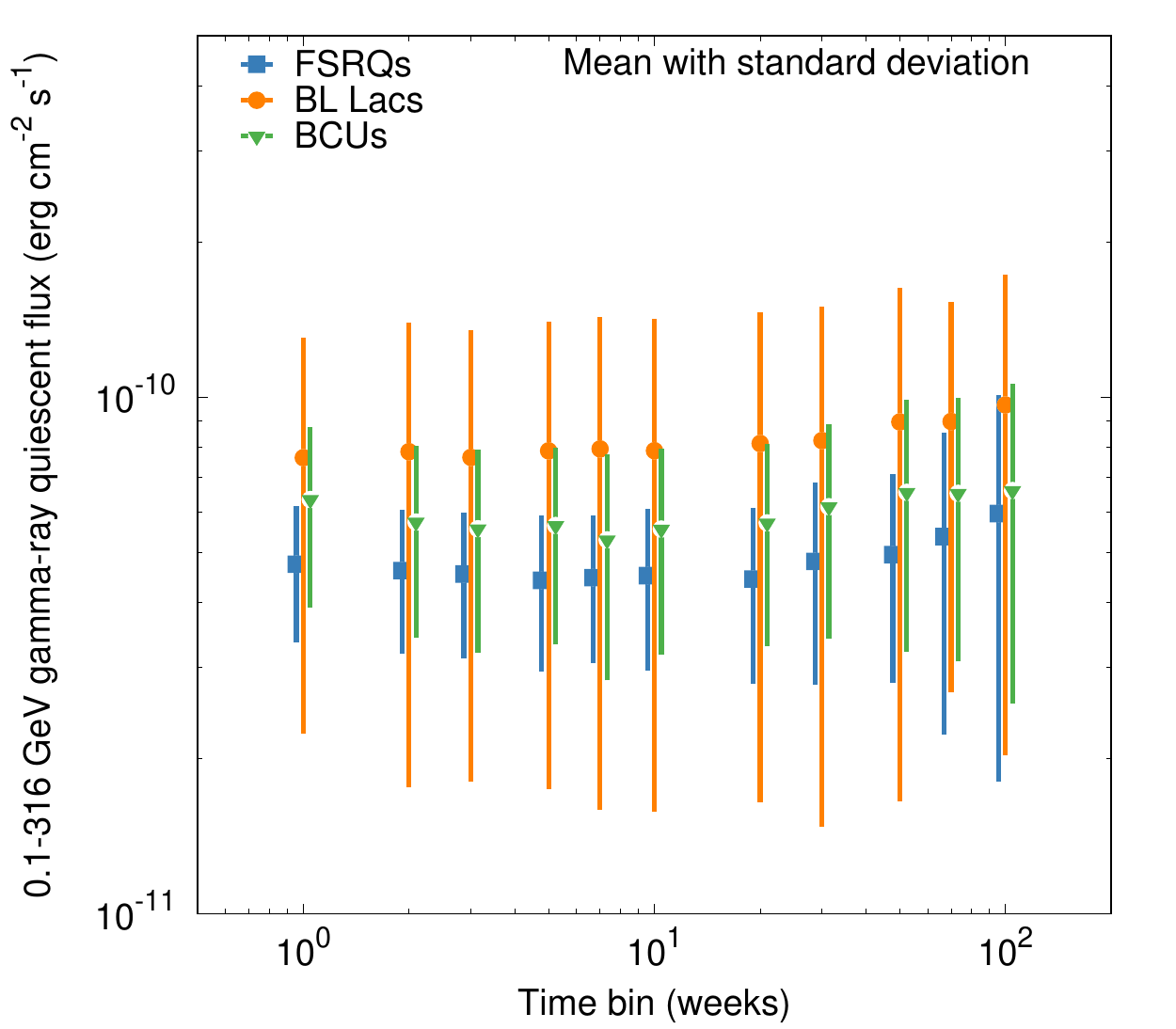}
\end{center}
\caption{The mean gamma-ray quiescent fluxes of 106 FSRQs (blue squares), 31 BL Lacs (orange circles), and 8 BCUs (green triangles) with the standard deviations as a function of time bins.}
\label{fig:binN2qbr}
\end{figure*}

\section{X-ray quiescent flux level} \label{sec:xray_quiescent_flux}

Examples of 1.0~keV {\swift} X-ray light curves are shown in  Fig.~\ref{fig:xray_quiescent} (black points). The Bayesian blocks representations are overplotted with solid orange lines. The quiescent flux levels are also presented with dashed cyan lines  and are estimated in the same way as for the gamma-ray quiescent flux levels  -- see Section \ref{sec:quiescent}.

\setcounter{figure}{0}
\begin{figure*}[h]
\gridline{
\fig{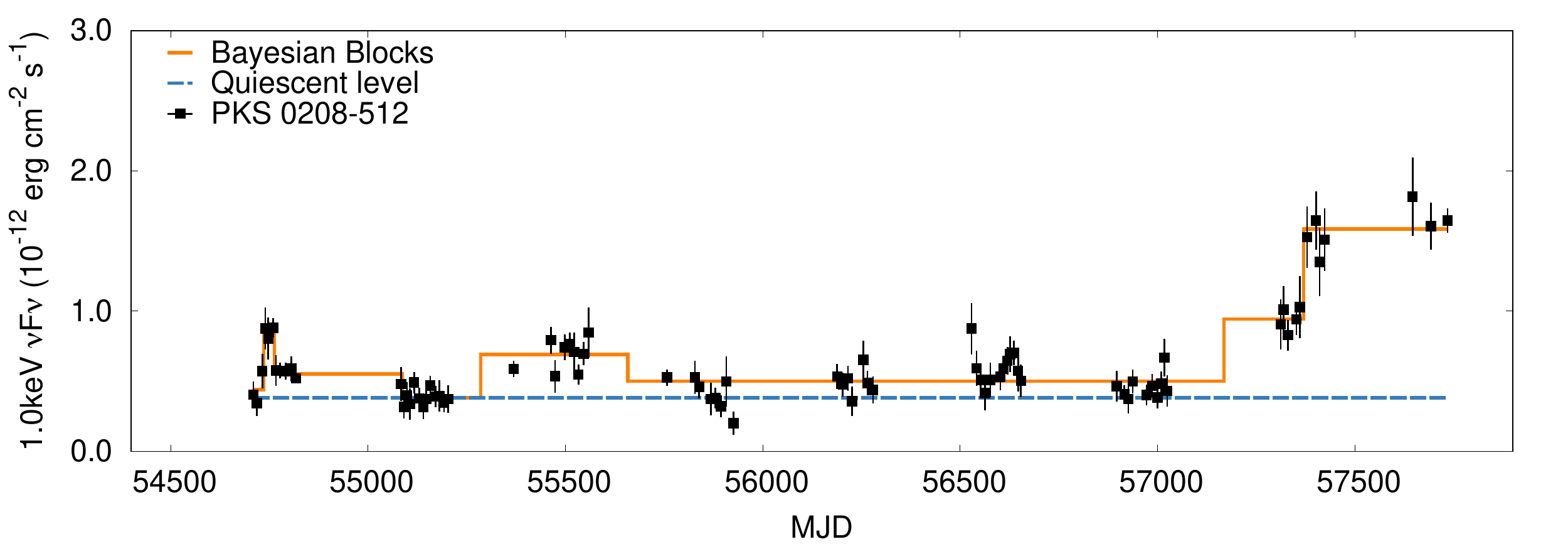}{0.49\textwidth}{(a) PKS 0208-512}
\fig{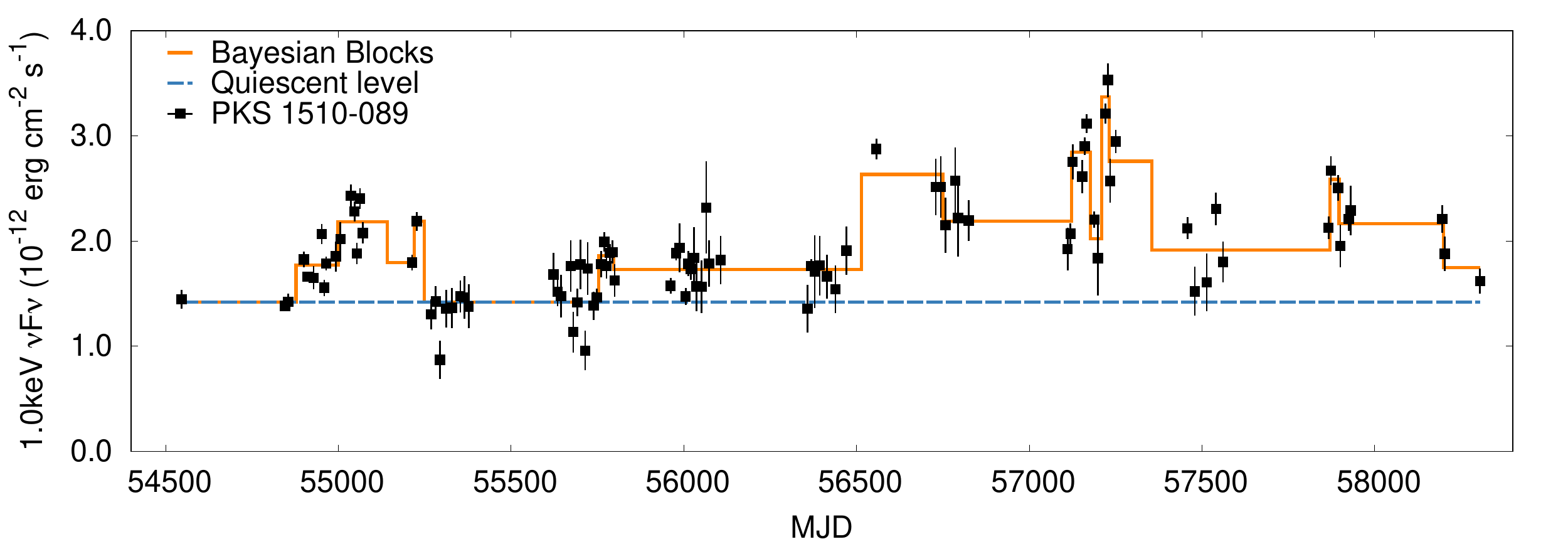}{0.49\textwidth}{(b) PKS 1510-089}
}
\caption{
Examples of the 1.0~keV quiescent fluxes (dashed cyan lines) of PKS 0208-512 and PKS 1510-089 X-ray light curves (black points) with the Bayesian block representations (orange solid lines).}
\label{fig:xray_quiescent}
\end{figure*}

\section{Comparison with photon flux distribution obtained by the FAVA analysis} \label{sec:fava}
The photon flux distribution has been used to study the time variation to hold some of the important clues to the origin and nature of their variability. 
\cite{Murase:2018} used the flux distributions with a convolution of a power law with a Poisson distribution, selecting 6 BL Lacs at intermediate redshifts from the FAVA analysis. Their analysis indicates that the power-law index $\alpha$ ranges from 1.7 to 3.0. However, it has been known that the FAVA analysis is not ideal for obtaining detailed information on gamma-ray light curves. Smaller values of $\alpha$ imply the larger output of neutrinos during the flaring phase, so it is important to determine $f_{\rm fl}$ and $b_{\rm fl}^\gamma$ more precisely.

In the standard leptonic scenario for the blazar gamma-ray emission, in common, the leptonic models predict a neutrino luminosity $L_{\nu}$ is in proportion to a gamma-ray luminosity $L_{\gamma}$ to the power of $\gamma \sim 1.5-2.0$ (see \cite{Murase:2016}, and references therein), giving
\begin{equation}
L_{\nu}^{2} \frac{dN}{dL_{\nu}} \propto L_{\nu}^{\frac{\gamma - \alpha + 1}{\gamma}}, 
\end{equation}
which implies that the flaring contribution can be dominant for the larger $\gamma$ and smaller $\alpha$, e.g., $\gamma \geq 1.5$ and $\alpha \leq 2.5$ \citep{Murase:2018}. 
Thus, the index $\alpha$ is a good indicator of the flaring contribution to the neutrino output. 
In this section, we compare our results of the flux distributions with those from the FAVA analysis. 

Our results for several indicative sources are presented in Table~\ref{tab:results-short}.
The source flux distribution per week is modeled as a convolution of a power law with a Poisson distribution of the source photons plus the background photons. 
While the number of the source photons is related to the source flux with the effective area and the exposure time, the effective area multiplied by the exposure time changes one-week bin by one-week bin. 
The number of the background photons changes as well. Hence, we include their systematic uncertainties to the fitting errors as the sum in quadrature of the errors. 
Figure~\ref{fig:flux_distribution} shows an example of the gamma-ray flux distributions fitted with a power-law function convolved with a Poisson distribution including the background photons, where the fitting is carried out with an unbinned maximum likelihood method. Our results on $\alpha$, ranging from 1.2 to 2.4, are systematically smaller than those of FAVA analysis, which range from 1.7 to 3.0 \citep{Murase:2018}. Hence, the photon flux distribution from the FAVA analysis tends to underestimate the flaring contribution to the neutrino output of a blazar, and our results would better justify the ``flare dominance'' in neutrino emission, considered by \cite{Murase:2018}. 

\setcounter{figure}{0}
\begin{figure}[h]
\gridline{
\fig{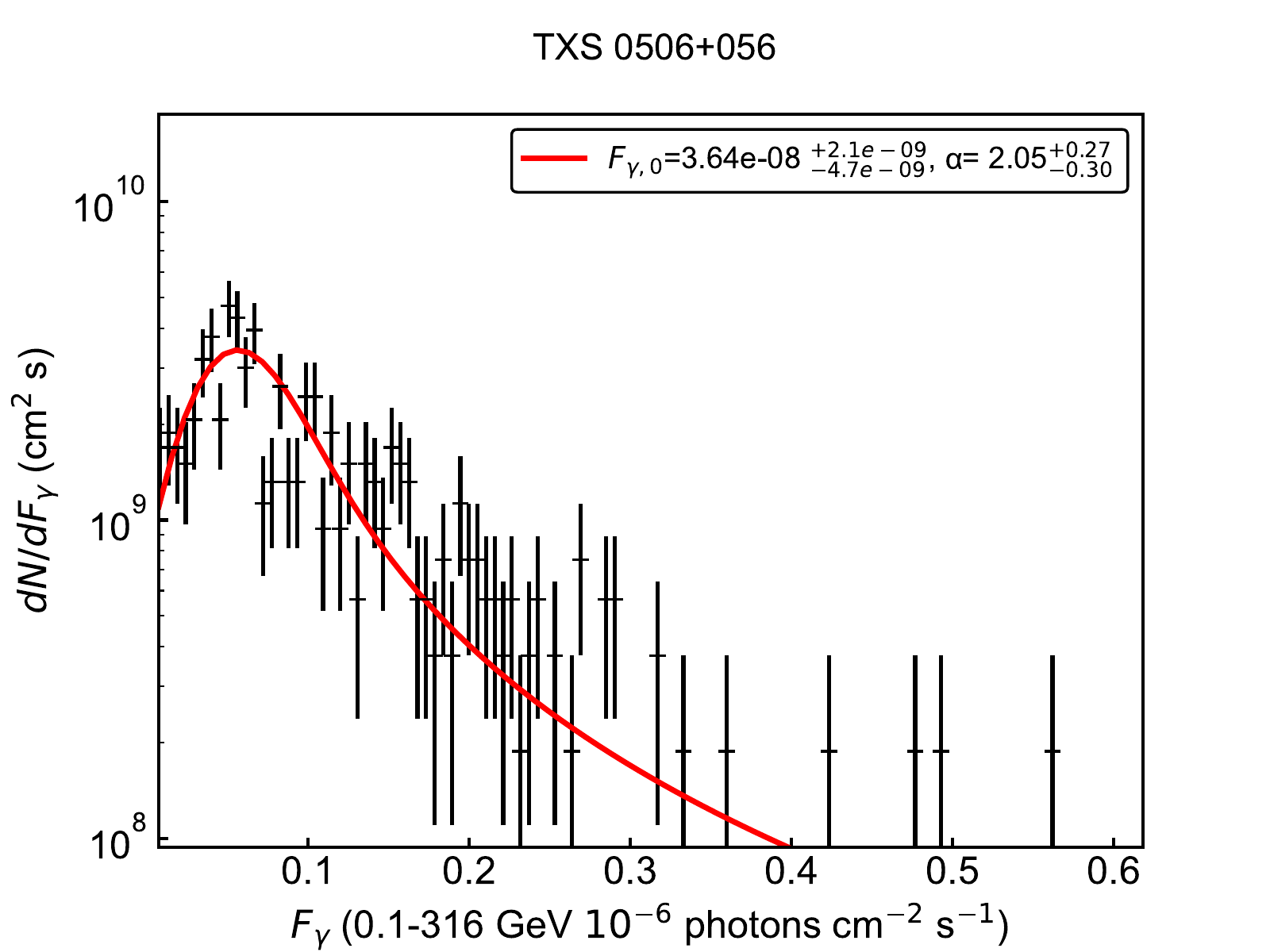}{0.45\textwidth}{}
\fig{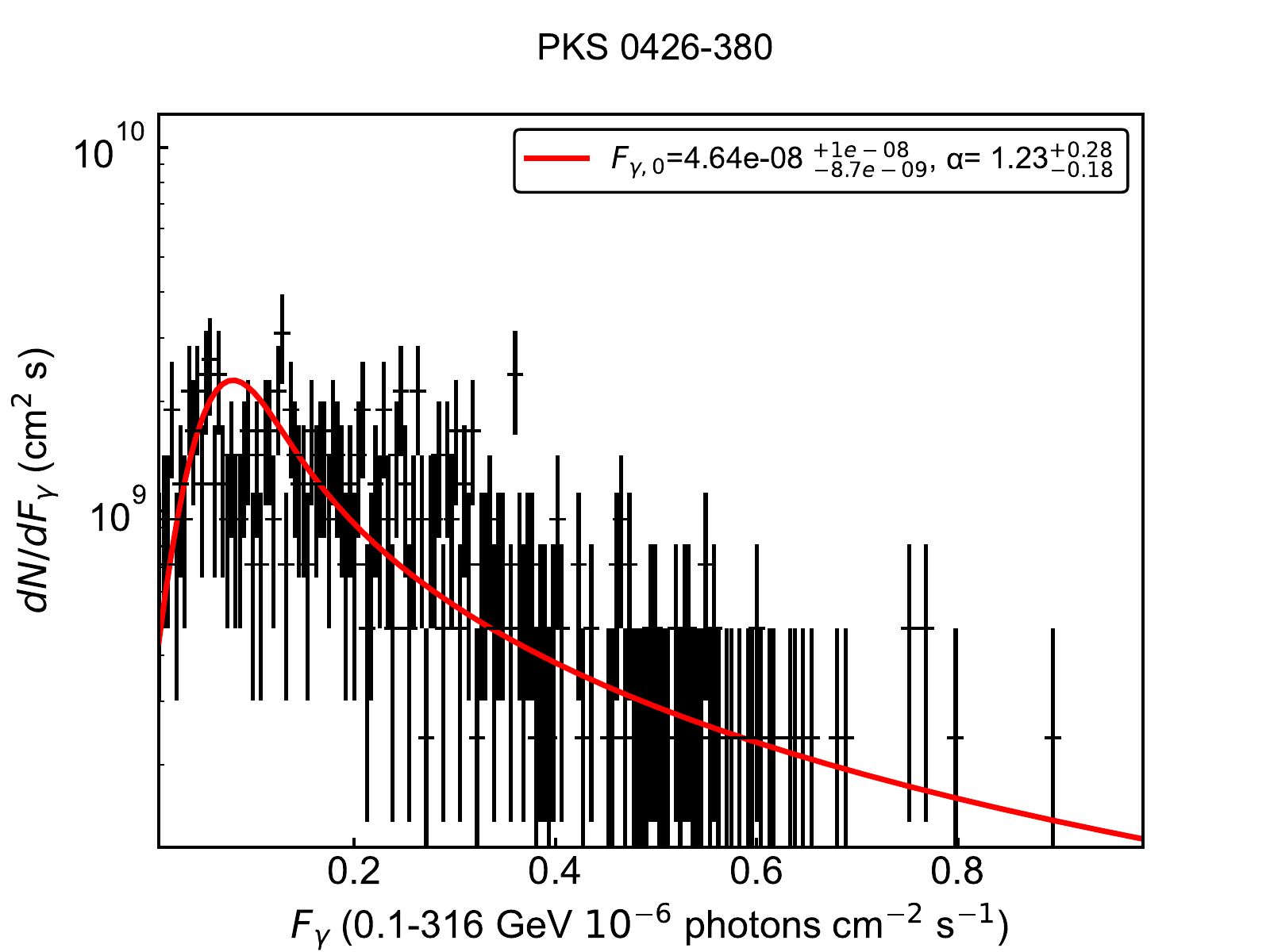}{0.45\textwidth}{}
}

\gridline{
\fig{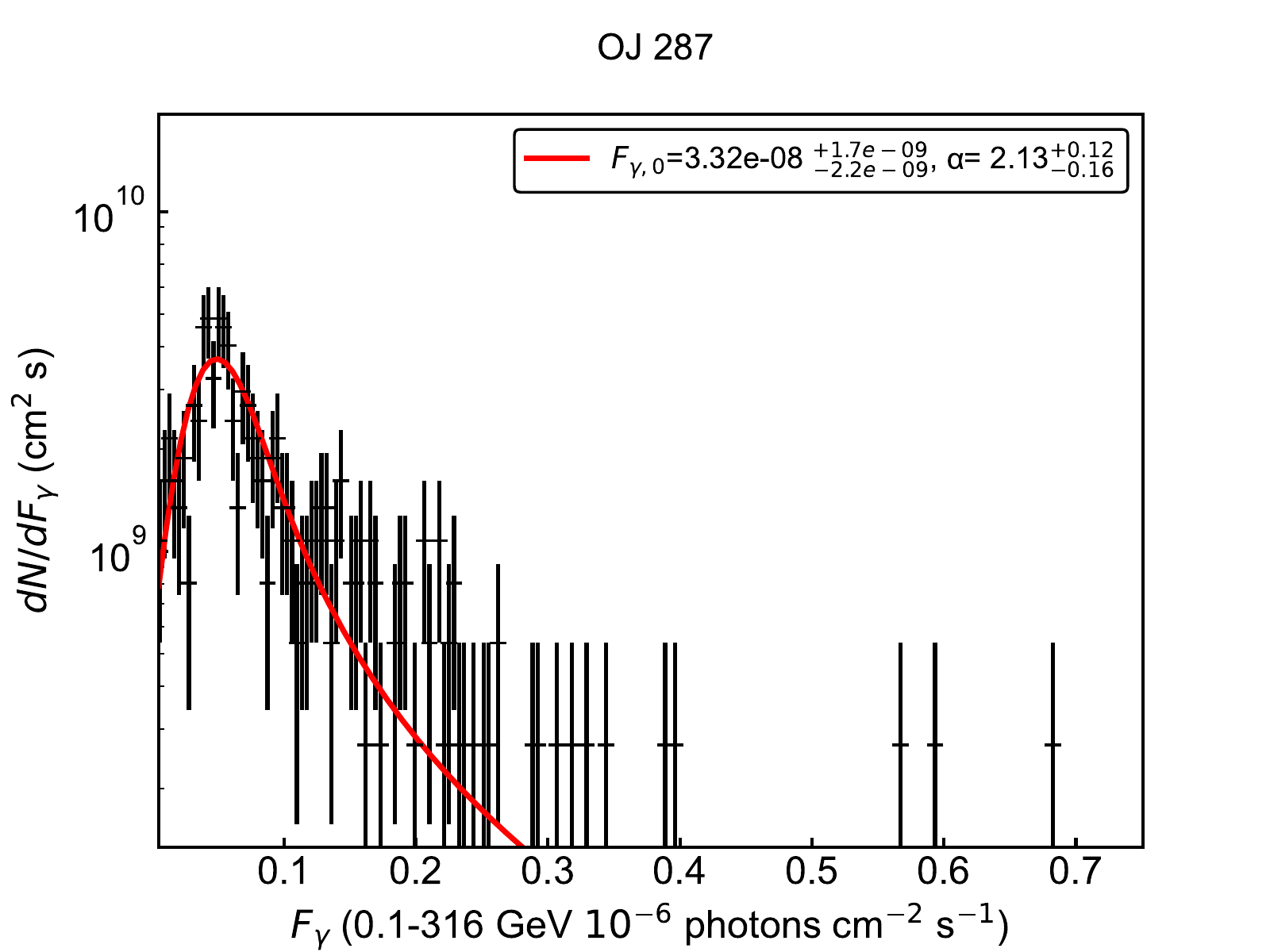}{0.45\textwidth}{}
\fig{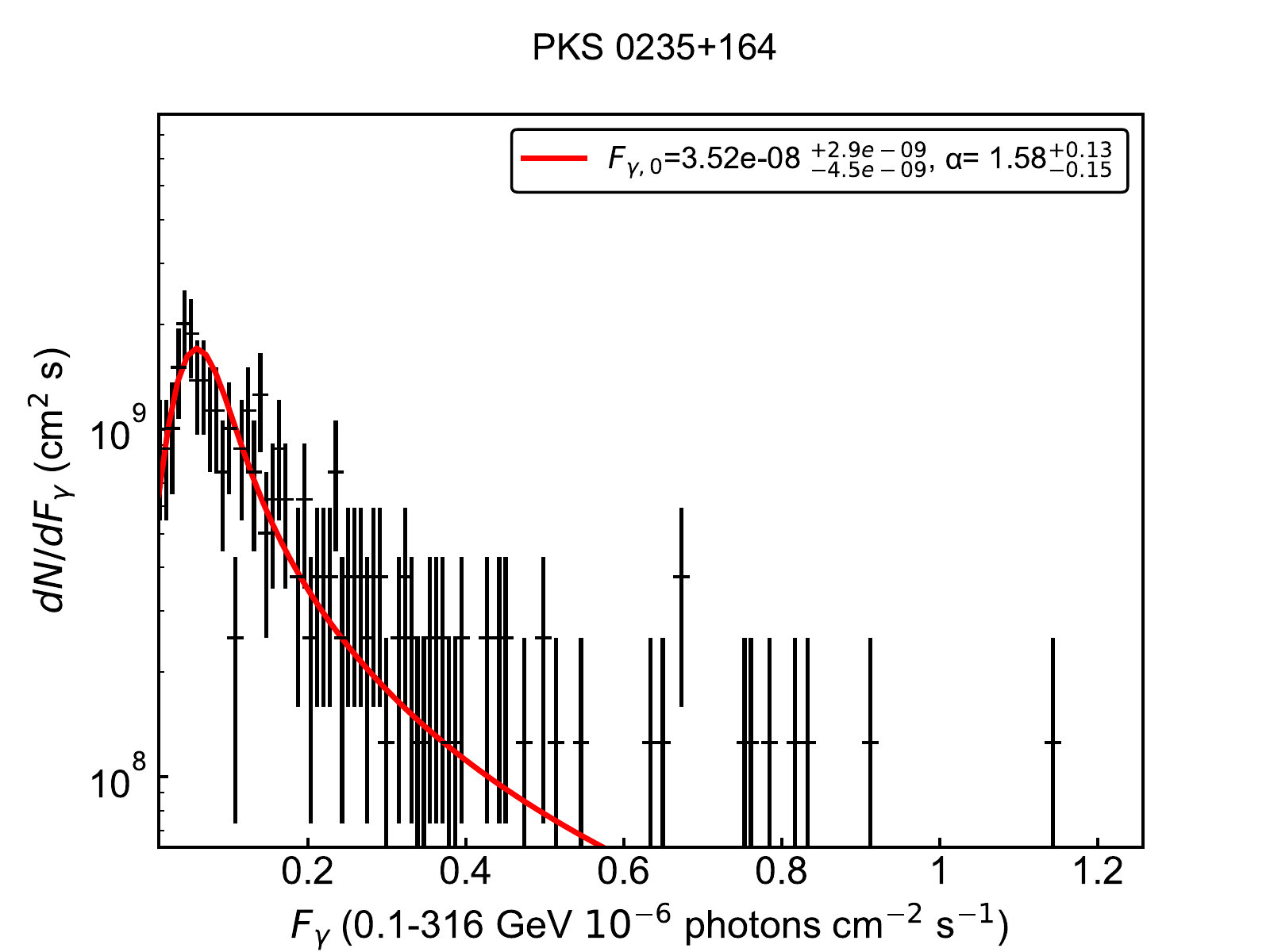}{0.45\textwidth}{}
}
\caption{
  The gamma-ray flux distributions of TXS 0506+056, PKS 0426-380, OJ 287, and PKS 0235+164 in 0.1--316~GeV fitted with a power-law function convolved with a Poisson distribution including the background photons. 
  See text for details.
}
\label{fig:flux_distribution}
\end{figure}

\setcounter{table}{0}
\begin{table*}[h]
    \caption{Gamma-ray flare duty cycle ($f_{\rm fl}$), flare energy fraction ($b^{\gamma}_{\rm fl}$), and best-fit power-law index of the flux distribution ($\alpha$) for a few indicative sources from our sample.}
\resizebox{0.99\linewidth}{!}{    
   \centering
    \begin{tabular}{c c c c c c c c c}
    \hline 
    \hline 
    Name & $f_{\rm fl}$ & $b^{\gamma}_{\rm fl}$ & $f^{\rm LE}_{\rm fl}$ &  $b^{\gamma{\rm LE}}_{\rm fl}$ &  $f^{\rm HE}_{\rm fl}$ & $b^{\gamma {\rm HE}}_{\rm fl}$ 
    	& $\alpha$ &  $\alpha^{\rm HE}$   \\
     \hline
TXS 0506+056 & $0.014 \pm 0.002$ & $0.055 \pm 0.011$ & $0.009 \pm 0.002$ & $0.039 \pm 0.009$ 
	& $0.038 \pm 0.004$ & $0.132 \pm 0.016$ & 
	$2.05_{-0.31}^{+0.27}$ &  $2.82_{-0.16}^{+0.15}$ \\
PKS 0426-380 & $0.164 \pm 0.008$ & $0.364 \pm 0.020$ & $0.136 \pm 0.008$ & $0.322 \pm 0.019$
	   & $0.152 \pm 0.008$ & $0.343 \pm 0.020$ & 
	   $1.23_{-0.18}^{+0.28}$ &  $2.18_{-0.20}^{+0.43}$ \\
OJ 287 & $0.031 \pm 0.004$ & $0.115 \pm 0.015$ & $0.029 \pm 0.004$ & $0.110 \pm 0.015$
   	& $0.023 \pm 0.003$ & $0.113 \pm 0.019$ & 
   	$2.13_{-0.16}^{+0.12}$  &  $2.84_{-0.19}^{+0.18}$ \\
PKS 0235+164 & $0.059 \pm 0.005$ & $0.183 \pm 0.017$ & $0.051 \pm 0.005$ & $0.165 \pm 0.017$
	   & $0.070 \pm 0.006$ & $0.212 \pm 0.019$ & 
	   $1.58_{-0.15}^{+0.13}$ & $1.87_{-0.09}^{+0.13}$ \\
PKS 0301-243 & $0.014 \pm 0.002$ & $0.076 \pm 0.016$ & $0.010 \pm 0.002$ & $0.067 \pm 0.016$
	   & $0.009 \pm 0.002$ & $0.058 \pm 0.017$ & 
	   $2.43_{-0.22}^{+0.21}$  & $3.15_{-0.16}^{+0.43}$ \\
S5 0716+71 & $0.290 \pm 0.011$ & $0.507 \pm 0.021$ & $0.221 \pm 0.010$ & $0.415 \pm 0.020$
	   & $0.224 \pm 0.010$ & $0.463 \pm 0.022$ & 
	   $1.43_{-0.13}^{+0.19}$ & $2.15_{-0.12}^{+0.13}$  \\
S4 0954+65 & $0.024 \pm 0.003$ & $0.114 \pm 0.018$ & $0.021 \pm 0.003$ & $0.104 \pm 0.018$
	   & $0.021 \pm 0.003$ & $0.138 \pm 0.024$ & 
	   $2.37_{-0.17}^{+0.13}$ & $2.51_{-0.17}^{+0.22}$ \\
     \hline
    \end{tabular}
}    
    \tablecomments{Values are reported for a low-energy range (LE: $0.1-1$ GeV), a high-energy range (HE: $1-316$ GeV), and the full LAT energy range ($0.1-316$ GeV). The flaring threshold level is defined in Equation (\ref{eq:threshold}) with $s=6$.}
    \label{tab:results-short}
\end{table*}

 \section{Estimated muon neutrino flare fluxes by using the gamma-ray photon fluxes.} \label{sec:neutrino_gphoton}

Figure \ref{fig:sindecl2nufl_gqb_ph_A1g15} shows the estimated muon neutrino flare fluxes from the gamma-ray photon fluxes for a one-week bin (a) and a 10-year bin (b) of scenario 1 with $A_{\gamma} = 1.0$ and ${\gamma} = 1.5$ as a function of ${\sin}({\delta})$. 
As seen by comparing Fig.~\ref{fig:sindecl2nufl_gqb_ph_A1g15} with Fig.~\ref{fig:sindecl2nufl_gqb_A1g15}, the estimated muon neutrino fluxes of the gamma-ray photon fluxes are almost similar to those of the gamma-ray energy fluxes.

\setcounter{figure}{0}
\begin{figure*}
\gridline{\fig{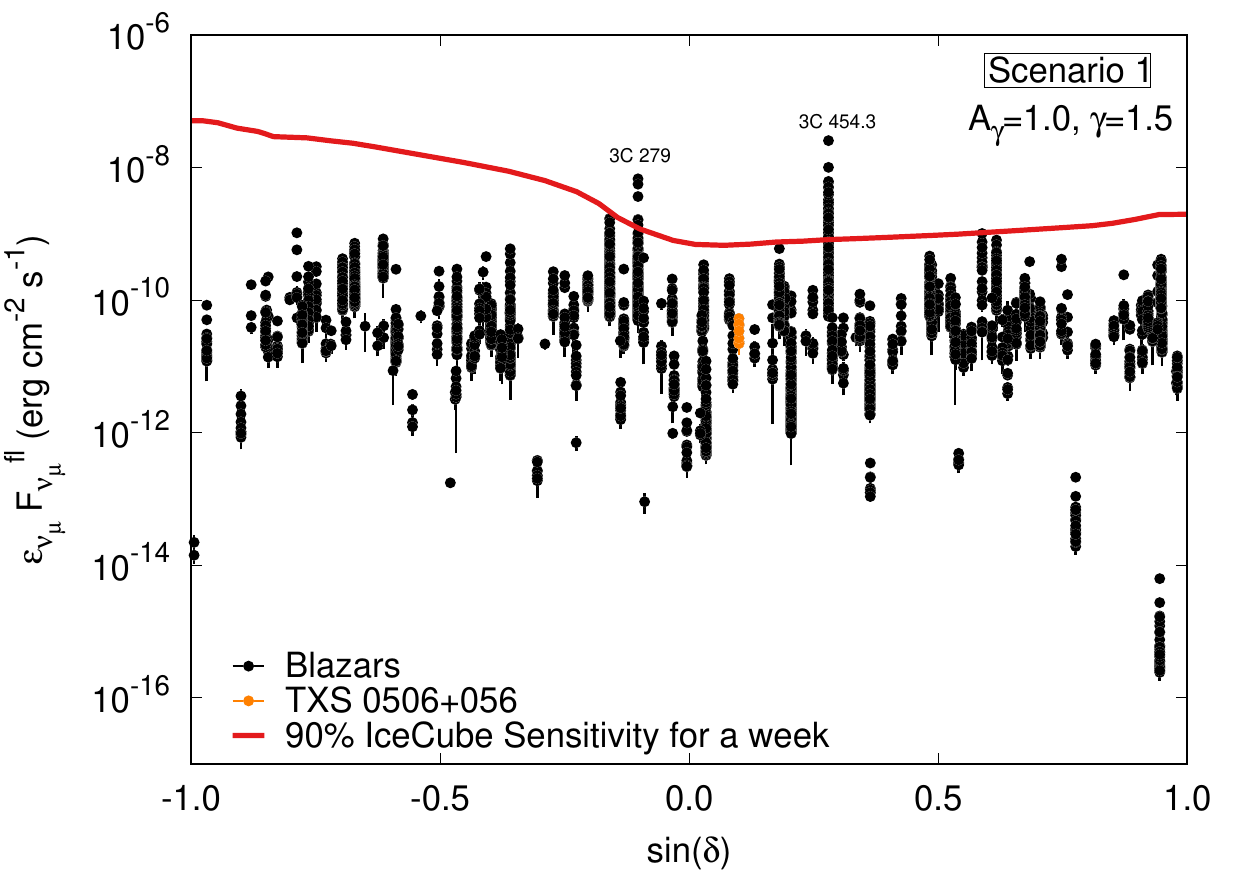}{0.48\textwidth}{(a)}
          \fig{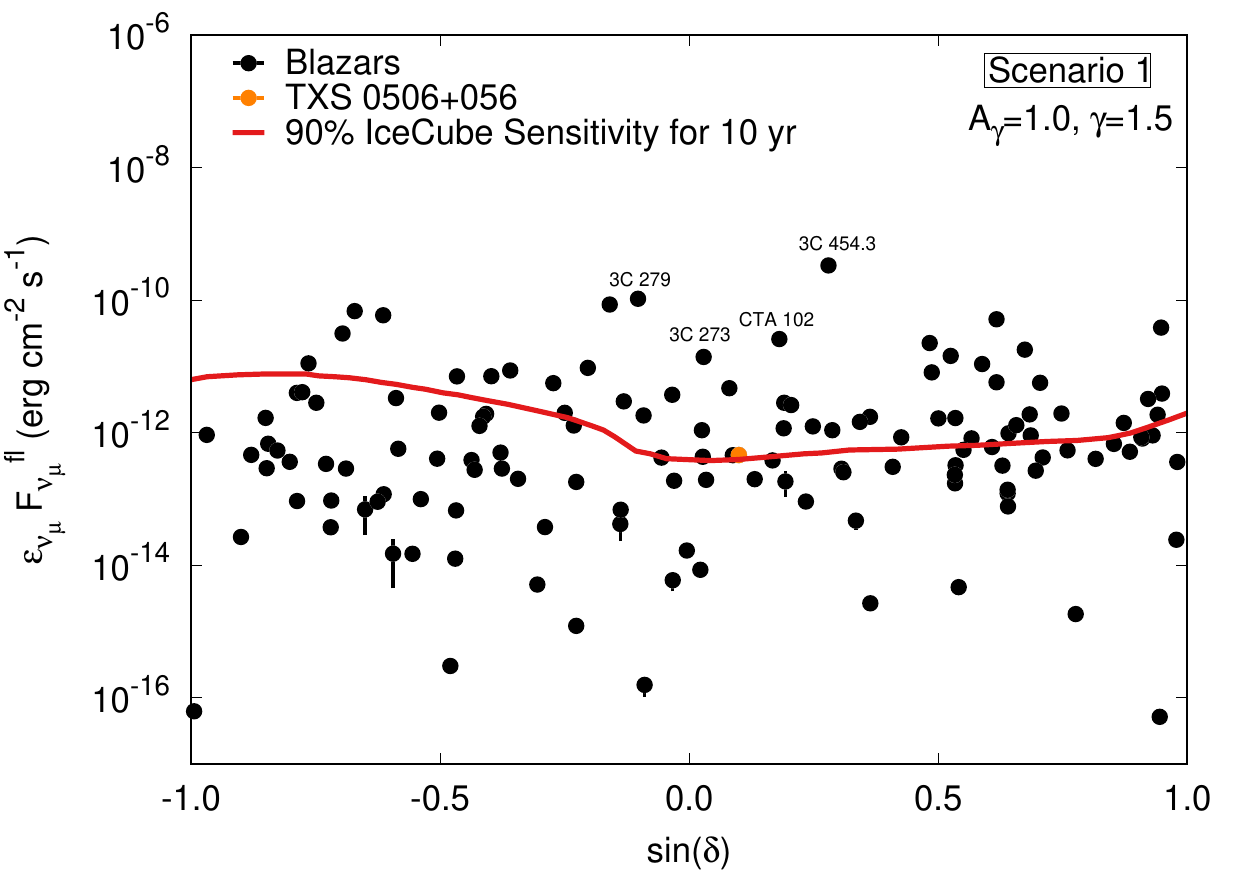}{0.48\textwidth}{(b)}
}
\caption{Estimated muon neutrino flare fluxes of 136 blazars (101 FSRQs, 28 BL Lacs, and 7 BCUs) from the gamma-ray photon fluxes for a one-week bin (a) and a 10-year bin (b) of scenario 1 with $A_{\gamma} = 1.0$ and ${\gamma} = 1.5$ as a function of ${\sin}({\delta})$. The 90~\% IceCube sensitivities are the same as in Figure \ref{fig:sindecl2nufl_gqb_A1g15}.}
\label{fig:sindecl2nufl_gqb_ph_A1g15}   
\end{figure*}



\end{document}